\documentclass[prb,twocolumn,aps,english,showpacs,superscriptaddress]{revtex4-1}

\usepackage{mathrsfs}
\usepackage{amsmath}
\usepackage{amssymb}
\usepackage{cancel}
\usepackage{pifont} 
\newcommand{\xmark}{\ding{55}}%
\usepackage{wasysym}
\setcitestyle{square}

\usepackage{graphicx}
\usepackage{url}
\usepackage{color}
\usepackage[colorlinks,citecolor=blue,linkcolor=blue,urlcolor=blue]{hyperref}

\begin{document}

\title{CT-X: an efficient continuous-time quantum Monte Carlo impurity solver \\
in Kondo Regime }

\author{Changming Yue}
\affiliation{Beijing National Laboratory for Condensed Matter Physics,
  and Institute of Physics, Chinese Academy of Sciences, Beijing
  100190, China}

\author{Yilin Wang}
\affiliation{Beijing National Laboratory for Condensed Matter Physics,
  and Institute of Physics, Chinese Academy of Sciences, Beijing
  100190, China}

\author{Junya Otsuki}
\affiliation{Department of Physics, Tohoku University, Sendai 980-8578, Japan}

\author{Xi Dai}
\email{daix@iphy.ac.cn}
\affiliation{Beijing National Laboratory for Condensed Matter Physics,
  and Institute of Physics, Chinese Academy of Sciences, Beijing
  100190, China}
\affiliation{Collaborative Innovation Center of Quantum Matter,
  Beijing, China}

\date{\today}

\begin{abstract}
In the present paper, we present an efficient continuous-time quantum Monte Carlo impurity
solver with high acceptance rate at low temperature for multi-orbital
quantum impurity models with general interaction. In this hybridization expansion
impurity solver, the imaginary time evolution operator for the high
energy multiplets, which decays very rapidly with the imaginary time,
is approximated by a probability normalized $\delta$-function. As the result, 
the virtual charge fluctuations of $f^{n}\rightarrow f^{n\pm1}$
are well included on the same footing without applying Schrieffer-Wolff
transformation explicitly. As benchmarks, our algorithm perfectly
reproduces the results for both Coqblin-Schriffeer and Kondo lattice
models obtained by CT-J method developed by Otsuki {\it et al}.
Furthermore, it allows capturing low energy physics of heavy-fermion
materials directly without fitting the exchange coupling $J$ in the
Kondo model. 
\end{abstract}
\maketitle

\section{Introduction}
 Due to the rapid development of hybridization expansion continuous-time
quantum Monte-Carlo (CT-HYB)\cite{rmp_ctqmc} method, an efficient 
solver for quantum impurity models, substantial progress has been achieved
in the electronic structure studies of strongly correlated materials within the framework of density functional
theory (DFT) implemented with dynamical mean-field theory (DMFT)\cite{rmp_dftdmft}. 
However, CT-HYB is insufficient for the studies of low temperature ($\sim$O(10)K)
properties of heavy fermion materials in the Kondo regime, where the
itinerant $s$, $p$, $d$ electrons co-exist and interact with the
localized $f$ electrons caused by the large Coulomb
repulsion $U$ among them. The failure of CT-HYB lies in its algorithm
construction where configurations with large charge fluctuations are
frequently proposed in the process of Monte Carlo updates, resulting in 
small acceptance rates in the Kondo regime where the charge fluctuations are nearly frozen. 
With the decrement of temperature {\it T}, CT-HYB method becomes increasingly inefficient because with the longer
imaginary time $\beta=1/k_{B}T$, configurations with large charge fluctuation are more and more
likely to be proposed during the sampling process, which has very small
acceptance rate.

One way to solve the above problem is to perform Schrieffer-Wolff transformation (SWT) \cite{SWTrans} to
single impurity Anderson model (SIAM) in the strong coupling limit and one gets effective low energy $s$-$d$
exchange model in which local charge fluctuations are projected out
and only virtual processes are considered. The well-known Coqblin-Shrieffer(CS)\cite{CSpaper}
model and Kondo model\cite{KMpaper} are two typical SW transformed models.
CT-J ~\cite{ctJ_paper} algorithm is developed to simulate such models by expanding partition
function in term of $s$-$d$ exchange terms. With much higher efficiency,
CT-J can be applied to study Kondo physics within the two localized
models down to much lower temperature. Based on the corresponding Kondo lattice model,
Matsumoto {\it et al}. have performed DMFT calculations for Ce-122 compounds
and successfully reproduced the general trend of antiferromagnetic transition temperature around the magnetic quantum
critical point\cite{KLM_prl2009}. In their approach, they first calculated hybridization function
between the conduction bands and the 4$f$ electrons by DFT+DMFT
with Hubbard-I approximation as an impurity solver and then constructed the effective CS model afterward by estimating $s$-$d$
exchange parameter $J$ obtained by SWT. However, such construction process 
neglects the fact that $J$ has momentum and orbital dependence.
Furthermore, once the realistic interactions (not the density-density type) among the $f$-electrons have been considered,
the SWT will become enormously tedious and complicated\cite{SW_f2S1_uranium}. 
As a result, CT-J is not the best practical choice for the calculations of the realistic heavy fermion materials.

In CT-HYB, the local trace part in the partition function can be viewed
as contributions from the various ``evolution paths''\cite{iqist} among different atomic
multiplets $\{\Gamma\}$ which can be grouped into high energy states
$\{\Gamma^{h}\}$ and low energy states $\{\Gamma^{l}\}$ according
to their atomic eigenenergy $E_{\Gamma}$. In Kondo regime, it is assumed
that $\{\Gamma^{l}\}$ are configurations with occupancy $n$, and $\{\Gamma^{h}\}$
are of occupancy $n\pm1$ with $n$ being an non-zero integer.
Furthermore, it is also assumed that $E_{\Gamma^{h}}\gg E_{\Gamma^{l}}$ as schematically 
shown in Fig.~\ref{ctx_approx}(a). In this condition, if one takes snapshots of the dynamics of electrons on the impurity site, 
atomic states would keep most time on low energy configurations for most of the time, as shown in Fig.~\ref{ctx_approx}(c).
The lower the energy is, the longer time it will spend on correspondingly. 
The imaginary-time evolution operator of the high energy
states, $e^{-E_{\Gamma^{h}}\tau}$, decays much faster than that of
$\Gamma^{l}$ as illustrated in Fig.~\ref{ctx_approx}(b). 
As $E_{\Gamma^{h}}$ increases,
the sharply decaying $e^{-E_{\Gamma^{h}}\tau}$ can be well approximated by the $\delta$-functions
centered at time zero, assuming that  $\Gamma^{h}$ appears only in the range of  $\tau\in[0,\sim0^{+})$.
Based on the above assumption, in the present paper we introduce a new impurity
solver by approximating $e^{-E_{\Gamma^{h}}\tau}$ with a probability
normalized $\delta$-function. With this new method, we are able to take into account all the virtual processes
 that involve the charge fluctuations from $\Gamma_l$ to $\Gamma_h$ states without explicitly applying SWT 
 which is difficult for the realistic materials. Furthermore, the approximation does
not depend on the details of local interaction and thus can be easily used for the DMFT 
calculations of the heavy fermion materials.

The rest of the paper is organized as follows. In the second section, we first
summarize the CT-HYB method and then introduce the
cutoff of the local Hilbert space. After that, we propose our
approximations to the local trace part in the partition functions for quantum impurity models under the Kondo limit. 
In Section III, we introduce how to design Monte-Carlo updates to sample  the partition
functions under the approximation mentioned above for both general and density-density type interactions.
Finally, the benchmarks of our new impurity solver are
shown in section IV on both CS and Kondo models. The summary of the paper is then given in section
V.

\section{Method}
\subsection{Hybridization Expansion}
Let us begin with the multi-band single impurity Anderson model (SIAM), which  reads
\begin{equation}
H_{\text{SIAM}}=H_{\text{loc}}+H_{\text{hyb}}+H_{\text{bath}},
\end{equation}
where
\begin{equation}
H_{\text{loc}}=\sum_{\alpha\beta}\epsilon_{\alpha\beta}f_{\alpha}^{\dagger}f_{\beta}+\sum_{\alpha\beta\delta\gamma}U_{\alpha\beta\delta\gamma}f_{\alpha}^{\dagger}f_{\beta}^{\dagger}f_{\gamma}f_{\delta},
\end{equation}
\begin{equation}
H_{\text{hyb}}=\sum_{\boldsymbol{k}\nu\alpha}V_{\boldsymbol{k}\nu}^{\alpha}c_{\boldsymbol{k}\nu}^{\dagger}f_{\alpha}+h.c.,
\end{equation}
\begin{equation}
H_{\text{bath}}=\sum_{\boldsymbol{k}\nu}\varepsilon_{\boldsymbol{k}\nu}c_{\boldsymbol{k}\nu}^{\dagger}c_{\boldsymbol{k}\nu}.
\end{equation}
The Greek letters $\alpha,\beta,\delta,\gamma$ denote $N_{0}$ localized
spin-orbital index, and $p\equiv\boldsymbol{k}\nu$ denotes the conduction
band (bath) electron with momentum $\boldsymbol{k}$ and spin-orbital
index $\nu$. 

The configuration space of hybridization expansion algorithm is given
by the set of imaginary times $\{\tau\}$ and corresponding orbital indices $\{\alpha\}$:
\begin{equation}
\mathcal{C}=\{\{\tau_{1},\cdots,\tau_{k}^{\prime};f_{\alpha_{1}},\cdots,f_{\alpha_{k}}^{\dagger}\}|k=0,1,\cdots\}
\label{Ck_original}
\end{equation}

Integrating out the bath operators $c_{p}(\tau)$, the partition function $Z$ reads (detailed derivations are given in Appendix A and Ref.~[\onlinecite{rmp_ctqmc}])
\begin{equation}
\begin{aligned}Z & =Z_{\text{bath}}\sum_{k=0}^{\infty}\int_{0}^{\beta}d\tau_{1}\cdots\int_{\tau_{k-1}}^{\beta}d\tau_{k}\int_{0}^{\beta}d\tau_{1}^{\prime}\cdots\int_{\tau_{k-1}^{\prime}}^{\beta}d\tau_{k}^{\prime}\\
 & \times\sum_{\alpha_{1}\cdots\alpha_{k}}\sum_{\alpha_{1}^{\prime}\cdots\alpha_{k}^{\prime}}w_{\text{loc}}(\mathcal{C}_{k})w_{\det}(\mathcal{C}_{k}),
\end{aligned}
\label{Z_original}
\end{equation}
where $w_{\text{loc}}(\mathcal{C}_{k})$ is the so-called local trace part
\begin{equation}
w_{\text{loc}}(\mathcal{C}_{k})=\text{Tr}_f[\mathcal{T}_{\tau}e^{-\beta H_{\text{loc}}}f_{\alpha_{k}}(\tau_{k})f_{\alpha_{k}^{\prime}}^{\dagger}(\tau_{k}^{\prime})\cdots f_{\alpha_{1}}(\tau_{1})f_{\alpha_{1}^{\prime}}^{\dagger}(\tau_{1}^{\prime})],
\end{equation}
 and $w_{\det}(\mathcal{C}_{k}$) is the so-called determinant part
\begin{equation}
w_{\det}(\mathcal{C}_{k})=\det\Delta^{(\mathcal{C}_{k})}.
\end{equation}
$\Delta^{(\mathcal{C}_k)}$ is a $k\times k$ matrix with its elements being anti-periodic hybridization
functions $\Delta^{(\mathcal{C}_{k})}_{ij}=\Delta_{\alpha_{i}\alpha_{j}^{\prime}}(\tau_{i}-\tau_{j}^{\prime})$,
\begin{equation}
\Delta_{\alpha_{i}\alpha_{j}^{\prime}}(\tau)=\sum_{p}\frac{V_{p}^{\alpha_{i}}V_{p}^{\alpha_{j}^{\prime}*}}{1+e^{-\varepsilon_{p}\beta}}\times\begin{cases}
e^{\varepsilon_{p}(\tau-\beta)}, & 0<\tau<\beta\\
-e^{\varepsilon_{p}\tau}, & -\beta<\tau<0
\end{cases},
\end{equation}
here $\tau\equiv\tau_{i}-\tau_{j}^{\prime}$. $\Delta$ can be reduced to a block-diagonal
matrix if the coupling to the bath is diagonal in spin-orbital indices,
and in this case we have $\det\Delta^{(\mathcal{C}_{k})}=\prod_{\alpha}\det\Delta^{(\mathcal{C}_{k})}_{\alpha}$.
We make this assumption of diagonal hybridization throughout the rest
of this paper. In practice, the inverse of $\Delta^{(\mathcal{C}_{k})}_{\alpha}$ denoted
by $\mathcal{M}^{(\mathcal{C}_{k})}_\alpha=[\Delta_\alpha^{(\mathcal{C}_{k})}]^{-1}$ is more convenient
to be saved and used in the fast-update formula\cite{fastupdate}.

When the interaction among $f$-electrons is density-density
type, the $w_{\text{loc}}(\mathcal{C}_{k})$ can be easily evaluated by segment
algorithm\cite{ctqmc_prl2006}. When the interaction term is the generalized type,
 the local Hamiltonian $H_{loc}$ is more complicated and the atomic eigenstates are no longer
 Fock states. In this case, the evaluation of the local trace part $w_{\text{loc}}(\mathcal{C}_{k})$ 
 becomes very time consuming and can be expressed in terms of the atomic eigenstates as
\begin{equation}
\begin{aligned}\omega_{\text{loc}}(\mathcal{C}_{k}) & =s_{T_{\tau}}\cdot\sum_{\Gamma_{1}\Gamma_{2}\cdots\Gamma_{2k}}e^{-(\beta-\tau_{k})E_{\Gamma_{1}}}\langle\Gamma_{1}|f_{\alpha_{k}}|\Gamma_{2k}\rangle\\
 & \times e^{-(\tau_{k}-\tau_{k}^{\prime})E_{\Gamma_{2k}}}\langle\Gamma_{2k}|f_{\alpha_{k}^{\prime}}^{\dagger}|\Gamma_{2k-1}\rangle\cdots\langle\Gamma_{3}|f_{\alpha_{1}^{\prime}}^{\dagger}|\Gamma_{2}\rangle\\
 & \times e^{-(\tau_{1}^{\prime}-\tau_{1})E_{\Gamma_{2}}}\langle\Gamma_{2}|f_{\alpha_{1}}|\Gamma_{1}\rangle e^{-(\tau_{1}-0)E_{\Gamma_{1}}},
\end{aligned}
\label{wloc_noaprox}
\end{equation}
where $s_{T_{\tau}}$ is the sign determined by the time-ordering of the fermionic operators.
Each term in Eq.~(\ref{wloc_noaprox}) can be diagrammatically illustrated as an evolution path\cite{iqist} of $\{\Gamma\}$, 
e.g. 
\begin{equation}
\beta\vdash\Gamma_{1}\xleftarrow{f_{\alpha_{k}}(\tau_{k})}\Gamma_{2k}\xleftarrow{f_{\alpha_{k}^{\prime}}^{\dagger}(\tau_{k}^{\prime})}\cdots\xleftarrow{f_{\alpha_{1}^{\prime}}^{\dagger}(\tau_{1}^{\prime})}\Gamma_{2}\xleftarrow{f_{\alpha_{1}}(\tau_{1})}\Gamma_{1}\dashv0,
\end{equation}
 which means that the local configuration evolves from $\Gamma_{1}$ at $\tau=0$
to other multiplets successively by annihilation or creation of electrons and finally
returns back to $\Gamma_{1}$ at $\tau=\beta$.

\subsection{Truncation of the Hilbert space }

For the sake of simplicity, $\{\Gamma\}$ can be divided into two classes, high energy
states $\{\Gamma^{h}\}$ and low energy states $\{\Gamma^{l}\}$.
In the Kondo limit, the average occupation number for the $f$-orbitals is very close to an integer, $n$,
which naturally defines the low energy atomic states with $n_f=n$.
The rest of the atomic states have much higher charging energy about several times of Hubbard U in difference
comparing to the low energy states. In the CS transformation, these high energy atomic states are treated 
as virtual processes, which lead to exchange interaction between the localized f-electrons and itinerant electrons 
in the $s,p,d$ bands. For instance, in Cerium compounds low energy states are $n_f=1$, and both the $n_f=0$ and $n_f=2$ states
are treated as virtual processes. Therefore for general SIAM with strong Coulomb
repulsion and deep local impurity level, the states $\{\Gamma^{h}\}=\{f^{n\pm1}\}$
are included as the virtual states. Now after the first step the local Hilbert space considered in our approach has been truncated to 
\begin{equation}
\{\Gamma\}=\{\Gamma^{l}|N_{\Gamma^{l}}=n\}\bigcup\{\Gamma^{h}|N_{\Gamma^{h}}=n\pm1\}.
\end{equation}
$\{\Gamma^{h}\}$ are still rarely visited in MC sampling whose energy
difference to $\{\Gamma^{l}\}$ is about several eV, which is one or two orders of magnitude larger than the typical Kondo temperature. 
In other words, the time evolution function, which determines the appearance probability of specific
atomic configurations in the MC processes, satisfies $e^{-\tau E_{\Gamma^{h}}}\ll e^{-\tau E_{\Gamma^{l}}}$
especially at low temperature. When $H_{loc}$ is in density-density form and segment picture is adopted, 
this implies the overlapping segments or anti-segments are very short. 

\begin{figure}[htp]
\includegraphics[clip,width=3.4in,angle=0]{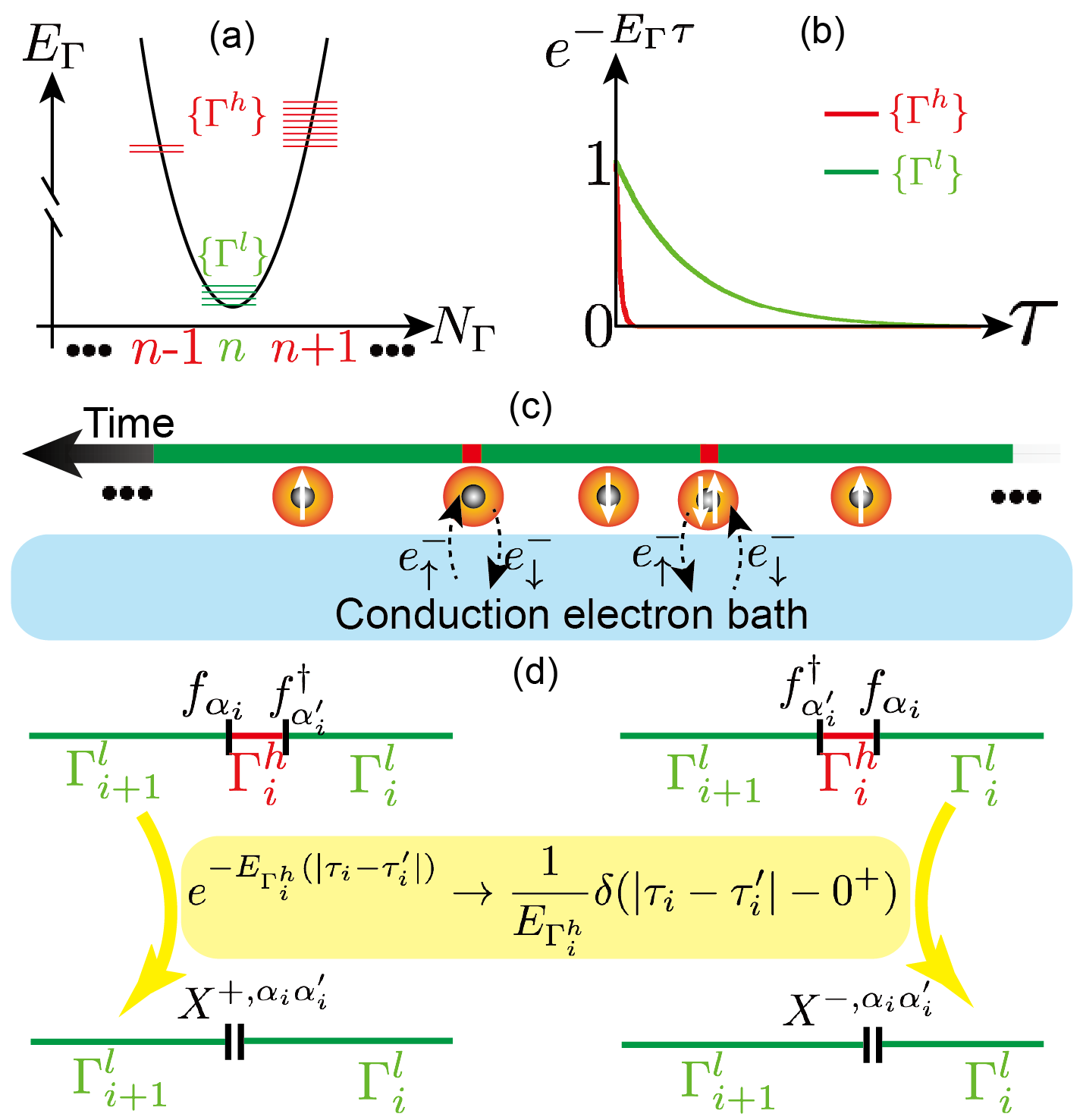}
\caption{(Color online). Approximations made to CT-HYB in Kondo regime. (a) Energy of atomic multiplets 
$\Gamma$ as a function of occupation number, in Kondo regime there is  $E_\Gamma^h \gg E_\Gamma^l$. 
(b) Schematic plot of $e^{-\Gamma\tau}$ as a function of $\tau$.  $e^{-\Gamma^h\tau}$ decays much faster than 
 $e^{-\Gamma^l\tau}$. (c) Schematic plot of the impurity site hybridizing with the heat bath.
In the simplest case of the single orbital model, impurity site spend most of the time on  
two low energy single occupied states, $|\uparrow\rangle$ and $|\downarrow\rangle$, than on two high energy states, unoccupied  $|\rangle$
and double occupied $|\uparrow\downarrow\rangle$ (adapted from Ref.~[\onlinecite{Gabriel_snapshot}]).  
(d) Sharply decaying imaginary time evolution operator of high energy states is approximated by a normalized Delta function, leading to virtual processes 
included in $X$ matrix. Left panel, $\tau_i>\tau_i^\prime$; right panel, $\tau_i<\tau_i^\prime$. }
\label{ctx_approx}
\end{figure}

The above truncation requires that evolution paths with non-zero contributions
to $w_{\text{loc}}(\mathcal{C}_{k})$ are those alternating $\{\Gamma^{h}\}$
and $\{\Gamma^{l}\}$ since
\begin{equation}
\begin{aligned}|\{\Gamma^{h}|N_{\Gamma^{h}}=n+1\}\rangle & \leftarrow f_{\alpha}^{\dagger}|\{\Gamma^{l}|N_{\Gamma^{l}}=n\}\rangle,\\
|\{\Gamma^{h}|N_{\Gamma^{h}}=n-1\} & \leftarrow f_{\alpha}|\{\Gamma^{l}|N_{\Gamma^{l}}=n\}\rangle,\\
|\emptyset\rangle & \leftarrow f_{\alpha}^{\dagger}|\{\Gamma^{h}|N_{\Gamma^{h}}=n+1\}\rangle,\\
|\emptyset\rangle & \leftarrow f_{\alpha}|\{\Gamma^{h}|N_{\Gamma^{h}}=n-1\}\rangle.
\end{aligned}
\end{equation}
 $w_{\text{loc}}(\mathcal{C}_{k})$ can be split into two parts according
to the energy hierarchy of the head/tail state $\{\Gamma_{1}\}$.
The part which starts from and ends in $\{\Gamma_{1}^{h}\}$ is generally
much smaller, since it contains more time evolution of the high energy
states and thus can be reasonably neglected, especially at very low 
temperature. Thus, we obtain
\begin{equation}
\begin{aligned}w_{\text{loc}}(\mathcal{C}_{k}) & \approx s_{T_{\tau}}\sum_{\Gamma_{1}^{l}\Gamma_{1}^{h}\cdots\Gamma_{k}^{l}\Gamma_{k}^{h}}e^{-(\beta-\tau_{k})E_{\Gamma_{1}^{l}}}\langle\Gamma_{1}^{l}|f_{\alpha_{k}}|\Gamma_{k}^{h}\rangle\\
 & \times e^{-(\tau_{k}-\tau_{k}^{\prime})E_{\Gamma_{k}}^{h}}\langle\Gamma_{k}^{h}|f_{\alpha_{k}^{\prime}}^{\dagger}|\Gamma_{k}^{l}\rangle\cdots\langle\Gamma_{2}^{l}|f_{\alpha_{1}^{\prime}}^{\dagger}|\Gamma_{1}^{h}\rangle\\
 & \times e^{-(\tau_{1}^{\prime}-\tau_{1})E_{\Gamma_{1}^{h}}}\langle\Gamma_{1}^{h}|f_{\alpha_{1}}|\Gamma_{1}^{l}\rangle e^{-(\tau_{1}-0)E_{\Gamma_{1}^{l}}},
\end{aligned}
\label{wloc_approx}
\end{equation}
which evolves in $\{\Gamma^{l}\}\leftarrow\{\Gamma^{h}\}\cdots\{\Gamma^{l}\}\leftarrow\{\Gamma^{h}\}\leftarrow\{\Gamma^{l}\}$.

\subsection{Energy shift}
Eigenvalues of $H_{\text{loc}}$, $\{E_{\Gamma}\}$, can be negative or positive,
therefore $e^{-\tau E_{\Gamma}}$($\tau>0$) is either monotonically increasing
or decreasing function, respectively. However, it is the relative difference between
$\{E_{\Gamma^h}\}$ and $\{E_{\Gamma_l}\}$  that matters in Monte Carlo simulations. 
Then it is convenient to make a shift to $\{E_{\Gamma}\}$
such that the time evolution functions appearing in our simulations are always monotonically decreasing. 
To realize that, we shift the zero of the energy to $E_0$, where  $E_{0}=\min\{E_{\Gamma^{l}}\}$,
\begin{equation}
E_{\Gamma}\rightarrow E_{\Gamma}^{\prime}=E_{\Gamma}-E_{0}\ge0.
\end{equation}
The transformation is equivalent to multiply a positive factor $e^{\beta E_{0}}$
to $w_{\text{loc}}(\mathcal{C}_{k})$ 
\begin{equation}
\omega_{\text{loc}}(\mathcal{C}_{k})\rightarrow\omega_{\text{loc}}^{\prime}(\mathcal{C}_{k})=\omega_{\text{loc}}(\mathcal{C}_{k})\times e^{\beta E_{0}},
\end{equation}
and partition function is changed to 
\begin{equation}
Z\rightarrow Z^{\prime}=Z\times e^{\beta E_{0}}.
\end{equation}
Please note that the expectation value of an operator will not be modified by the above transformation, 
\begin{equation}
\begin{aligned}\langle\hat{O}\rangle_{Z^{\prime}} & =\frac{\int d\mathcal{C}w^{\prime}(\mathcal{C})O(\mathcal{C})}{Z^{\prime}}=\frac{\int d\mathcal{C}w(\mathcal{C})e^{\beta E_{0}}O(\mathcal{C})}{Z\times e^{\beta E_{0}}}\\
 & =\frac{\int d\mathcal{C}w(\mathcal{C})O(\mathcal{C})}{Z}=\langle\hat{O}\rangle_{Z}.
\end{aligned}
\end{equation}
Prime $\prime$ is omitted for $E_{\Gamma}^{\prime}$, $Z^{\prime}$,
etc. hereafter for the sake of simplicity.

\subsection{Approximations in Kondo limit}

Two typical fragments of evolution paths appearing in $w_{\text{loc}}(\mathcal{C}_{k}$) in Eq.~(\ref{wloc_approx})
are schematically depicted in Fig.~\ref{ctx_approx}(d) where each high energy state is sandwiched 
between one creation and one annihilation operators.  Here we focus on the left panel where 
$\tau_{i}>\tau_{i}^{\prime}$. In the limit of $E_{\Gamma_{i}^{h}}\rightarrow+\infty$,
the probability of finding a configuration with finite $\tau=\tau_{i}-\tau_{i}^{\prime}>0$
approaches 0 due to the exponentially decreasing factor $e^{-(\tau_{i}-\tau_{i}^{\prime})E_{\Gamma_{i}^{h}}}$,
which means that excitations to high energy states are instantaneous, i.e. $\tau_{i}-\tau_{i}^{\prime}=0^{+}$, 
in the Kondo limit. Integrating over $\tau_{i}^{\prime}$, we
obtain 
\begin{equation}
\int_{\tau_{i-1}}^{\tau_{i}}e^{-(\tau_{i}-\tau_{i}^{\prime})E_{\Gamma_{i}^{h}}}d\tau_{i}^{\prime}\cdots=\cdots\frac{1}{E_{\Gamma_{i}^{h}}}\cdots|_{E_{\Gamma_{i}^{h}}\rightarrow+\infty},
\end{equation}
where $\frac{1}{E_{\Gamma_{i}^{h}}}$ indicates total probability for this particular type of virtual processes.
Then in the Kondo limit, where all the high energy local atomic states can be treated as the virtual processes, 
the sharply decreasing time evolution can be well approximated by a probability normalized delta function
\begin{equation}
e^{-(\tau_{i}-\tau_{i}^{\prime})E_{\Gamma_{i}^{h}}}\rightarrow\frac{1}{E_{\Gamma_{i}^{h}}}\delta(\tau_{i}-\tau_{i}^{\prime}-0^{+}).
\end{equation}
This approximation is getting better and better when the charging energy $E_{\Gamma^h}$ is approaching infinity, 
which is very suitable for the heavy fermion systems at the Kondo limit. The above approximation has the following advantages. 
1) By neglecting the time dependence of the local propagator for the high energy atomic states, the charge fluctuations to these 
high energy atomic states will  be treated as the virtual processes, which induce an effective exchange coupling among the 
conduction electrons and the low energy atomic states.   For simple model system, i.e. the single orbital Anderson impurity model, 
it can automatically obtain the exact same coupling terms as the SWT. 2) This approximation can be easily applied to more realistic 
models generated during the process of LDA+DMFT, the coupling terms between the f-electrons and conduction electrons have the 
momentum and orbital dependence, which make SWT very difficult.

\begin{figure}[htp]
\includegraphics[clip,width=3.4in,angle=0]{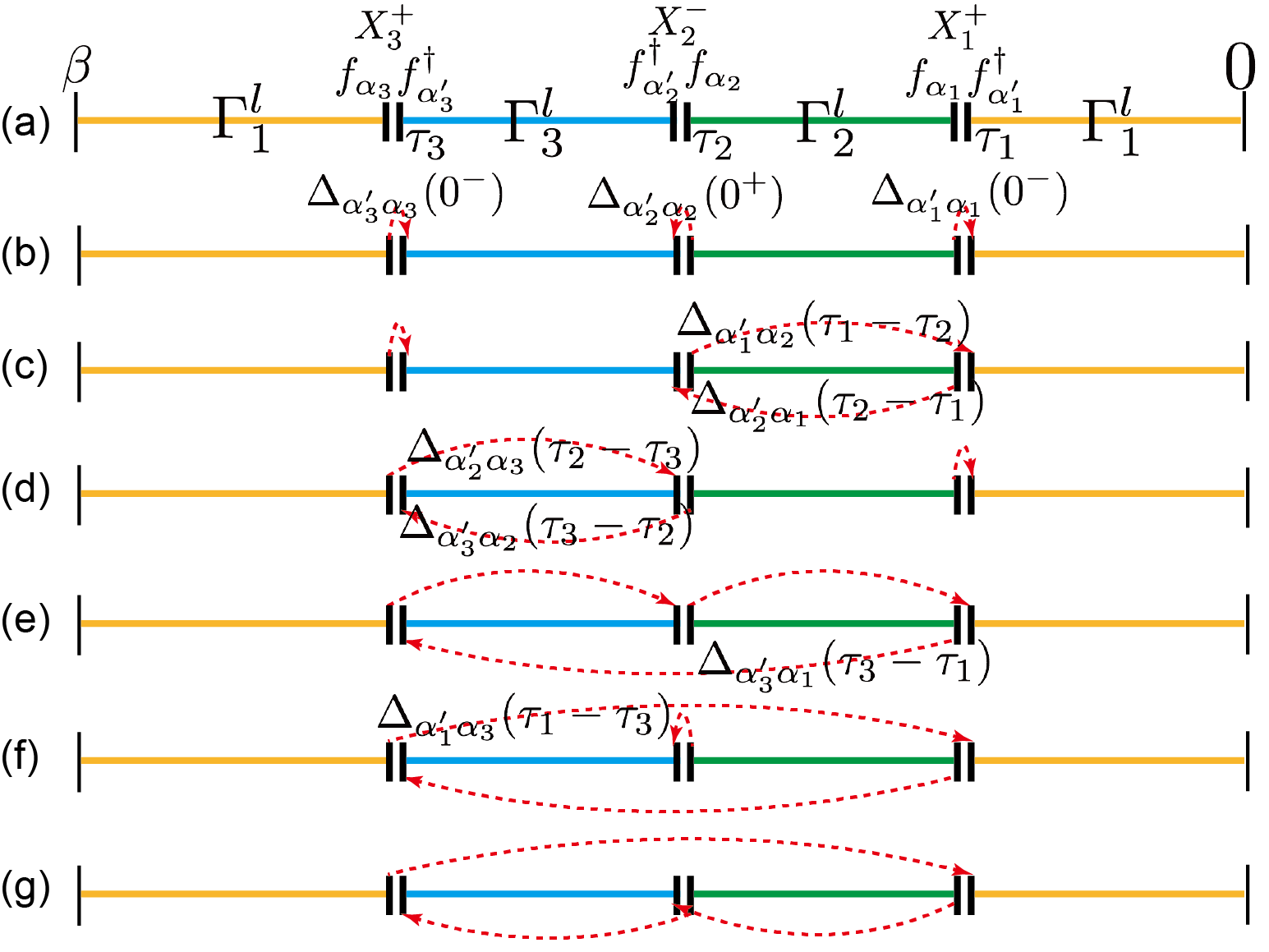}
\caption{(Color online) Schematic plot of a third order configuration in the approximated partition 
function Eq.~(\ref{zapprox}). (a), The evolution of low energy states ($\Gamma^l_i$, labelled by horizontal solid colored lines) by means of virtual processes.  Two adjacent vertical solid black lines are to denote the creation and annihilation operator pair in an $X$ matrix. $\pm$ in $X^{\pm}_i$ is to denote the type ($s_i=\pm1$) of $X$ matrices defined in Eq.~(\ref{xmatp1}, \ref{xmatm1}). (b-g), Illustration of the hybridizations determinant, Eq.~(\ref{detM}), by arrowed dashed red lines starting from a annihilation operator and ending at a creation operator. For a three order term, there are $3!=6$ terms in the determinant.} \label{diagram_c3}
\end{figure}

Replacing all $e^{-\tau\cdot E_{\Gamma^{h}}}$ with $\frac{1}{E_{\Gamma^{h}}}\delta(\tau-0^{+})$
and integrating over all $\{\tau_{i}^{\prime}\}$, we find that  a creation operator and an annihilation operator
always appear in adjacent pairs. The configuration space now reads 
\begin{equation}
\begin{aligned}\mathcal{C} & =\{\{\},\{\tau_{1};s_{1}f_{\alpha_{1}}f_{\alpha_{1}}^{\dagger}\},\cdots,\\
 & \times\{\tau_{1},\cdots,\tau_{k};\:s_{1}f_{\alpha_{1}}f_{\alpha_{1}^{\prime}}^{\dagger},\cdots,s_{k}f_{\alpha_{k}}f_{\alpha_{k}^{\prime}}^{\dagger}\},\cdots\},
\end{aligned}
\end{equation}
where 
\begin{equation}
\begin{aligned}s_{i}f_{\alpha_{i}}f_{\alpha_{i}^{\prime}}^{\dagger}|_{s_{i}=1} & \equiv f_{\alpha_{i}}f_{\alpha_{i}^{\prime}}^{\dagger}\rightarrow\tau_{i}=\tau_{i}^{\prime}+0^{+},\\
s_{i}f_{\alpha_{i}}f_{\alpha_{i}^{\prime}}^{\dagger}|_{s_{i}=-1} & \equiv f_{\alpha_{i}^{\prime}}^{\dagger}f_{\alpha_{i}}\rightarrow\tau_{i}=\tau_{i}^{\prime}-0^{+}.
\end{aligned}
\end{equation}
Summation over $\{\Gamma^{h}\}$ can be written in a compact form
by defining two types of $X$-matrices labelled by $s$ 
\begin{equation}
\begin{aligned}X_{\Gamma_{i+1}^{l}\Gamma_{i}^{l}}^{s_{i}\alpha_{i}\alpha_{i}^{\prime}}|_{s_{i}=1} & \equiv\sum_{\Gamma_{i}^{h}}\langle\Gamma_{i+1}^{l}|f_{\alpha_{i}}|\Gamma_{i}^{h}\rangle\frac{1}{E_{\Gamma_{i}^{h}}}\langle\Gamma_{i}^{h}|f_{\alpha_{i}^{\prime}}^{\dagger}|\Gamma_{i}^{l}\rangle,\end{aligned}
\label{xmatp1}
\end{equation}
 which describes virtual charge excitations $f^{n}\rightarrow f^{n+1}$ and 
\begin{equation}
\begin{aligned}X_{\Gamma_{i+1}^{l}\Gamma_{i}^{l}}^{s_{i}\alpha_{i}\alpha_{i}^{\prime}}|_{s_{i}=-1} & \equiv\sum_{\Gamma_{i}^{h}}\langle\Gamma_{i+1}^{l}|f_{\alpha_{i}^{\prime}}^{\dagger}|\Gamma_{i}^{h}\rangle\frac{1}{E_{\Gamma_{i}^{h}}}\langle\Gamma_{i}^{h}|f_{\alpha_{i}}|\Gamma_{i}^{l}\rangle,\end{aligned}
\label{xmatm1}
\end{equation}
which describes virtual charge excitations from $f^{n}\rightarrow f^{n-1}$
. Finally one obtains the partition function 
\begin{equation}
\begin{aligned}Z & \approx Z_{\text{bath}}\sum_{k=0}^{\infty}\int_{0}^{\beta}d\tau_{1}\cdots\int_{\tau_{k-1}}^{\beta}d\tau_{k}\sum_{\substack{\alpha_{1}\cdots\alpha_{k}\\
\alpha_{1}^{\prime}\cdots\alpha_{k}^{\prime}}} s_{T_{\tau}}\\
 & \times w_{\text{loc}}(\mathcal{C}_{k})\prod_{\alpha}\det(\mathcal{M}_\alpha^{(k_\alpha)})^{-1},
\end{aligned}
\label{zapprox}
\end{equation}
where the local trace is reformulated in terms of $X$-matrices as
\begin{equation}
\begin{aligned}w_{\text{loc}}(\mathcal{C}_{k}) & = \sum_{\Gamma_{1}^{l}\cdots\Gamma_{k}^{l}} e^{-(\beta-\tau_{k})E_{\Gamma_{1}^{l}}}X_{\Gamma_{1}^{l}\Gamma_{k}^{l}}^{s_{k}\alpha_{k}\alpha_{k}^{\prime}}e^{-(\tau_{k}-\tau_{k-1})E_{\Gamma_{k}^{l}}}\\
& \times \cdots X_{\Gamma_{3}^{l}\Gamma_{2}^{l}}^{s_{2}\alpha_{2}\alpha_{2}^{\prime}} e^{-(\tau_{2}-\tau_{1})E_{\Gamma_{2}^{l}}}X_{\Gamma_{2}^{l}\Gamma_{1}^{l}}^{s_{1}\alpha_{1}\alpha_{1}^{\prime}}e^{-(\tau_{1}-0)E_{\Gamma_{1}^{l}}}.
\end{aligned}
\label{wloc_ctx}
\end{equation}
An example of third order configuration $\mathcal{C}_{3}$ is schematically shown in Fig.~\ref{diagram_c3}(a) and its determinant part is 
\begin{widetext}
\begin{equation}
w_{\det}(\mathcal{C}_{3})=\left|\begin{array}{ccc}
\triangle_{\alpha_{1}^{\prime}\alpha_{1}}(0^{-}) & \triangle_{\alpha_{1}^{\prime}\alpha_{2}}(\tau_{1}-\tau_{2}) & \triangle_{\alpha_{1}^{\prime}\alpha_{3}}(\tau_{1}-\tau_{3})\\
\triangle_{\alpha_{2}^{\prime}\alpha_{1}}(\tau_{2}-\tau_{1}) & \triangle_{\alpha_{2}^{\prime}\alpha_{2}}(0^{+}) & \triangle_{\alpha_{2}^{\prime}\alpha_{3}}(\tau_{2}-\tau_{3})\\
\triangle_{\alpha_{3}^{\prime}\alpha_{1}}(\tau_{3}-\tau_{1}) & \triangle_{\alpha_{3}^{\prime}\alpha_{2}}(\tau_{3}-\tau_{2}) & \triangle_{\alpha_{3}^{\prime}\alpha_{3}}(0^{-})
\end{array}\right|,
\label{detM}
\end{equation}
\end{widetext}
which can be expanded into $3!=6$ terms as schematically represented in Fig.~\ref{diagram_c3}(b-g).

\section{Monte Carlo sampling}

Before introducing the detail of the Monte Carlo sampling, we first divide the different pair operators into the following types
\begin{itemize}
\item pure-pair: $\alpha_{i}=\alpha_{i}^{\prime}$, $sf_{\alpha_{i}}f_{\alpha_{i}}^{\dagger}$,
\item mix-pair: $\alpha_{i}\ne\alpha_{i}^{\prime}$, $s_{i}f_{\alpha_{i}}f_{\alpha_{i}^{\prime}}^{\dagger}$. 
\end{itemize}
An $k$-th order configuration $\mathcal{C}_{k}$ consists of time-ordered pure-pairs and mix-pairs 
\begin{equation}
\beta\vdash s_{k}f_{\alpha_{k}}f_{\alpha_{k}^{\prime}}^{\dagger}(\tau_{k})-\cdots-s_{1}f_{\alpha_{1}}f_{\alpha_{1}^{\prime}}^{\dagger}(\tau_{1})\dashv0.
\end{equation}
$\mathcal{C}_{k}$ contains an equal number of creation and annihilation
operators for each flavor by construction. With fixed $\{\tau_{i}\}$
and fixed number of single-particle operators of each flavor, $\mathcal{C}_{k}$
is mathematically an element in the set of direct products of operators
permutations $\mathscr{P}$,
\begin{equation}
\begin{aligned}\{\mathcal{C}_{ki}\} & =\{\mathscr{P}\{f_{\alpha_{k}},\cdots,f_{\alpha_{1}}\}\bigotimes\mathscr{P}\{f_{\alpha_{k}^{\prime}}^{\dagger},\cdots,f_{\alpha_{1}^{\prime}}^{\dagger}\}\\
 & \bigotimes\prod_{i=1}^{k}\mathscr{P}\{f_{\alpha_{i}},f_{\alpha_{i}^{\prime}}^{\dagger}\}\}.
\end{aligned}
\end{equation}
Based on the fact that any permutation can be expressed as the product
of transpositions, we design updates which keep diagram order as the
following,
\begin{itemize}
\item left-exchange: exchange annihilation operators of two adjacent pairs,
\item right-exchange: exchange creation operators of two adjacent pairs, 
\item in-pair swap: $s_{i}\rightarrow-s_{i}$.
\end{itemize}
Ergodicity can be satisfied by the above updates together with insertion
and removal of pure-pairs at random times which change expansion order
by 1, since any $\mathcal{C}_{ki}$ can be generated from an list
of pure pairs by successive transpositions. Updates which shift pair-operators
is not necessary but is useful to increase sampling efficiency. 

Metropolis-Hastings algorithm is used to sample configuration space
$\mathcal{C}$ according to the configuration weight 
$w(\mathcal{C}_{k})=w_{\text{loc}}(\mathcal{C}_{k})\prod_{\alpha}\det(\mathcal{M}_{\alpha}^{(\mathcal{C}_{k})})^{-1}d\tau^{k}$.
The random walk in $\mathcal{C}$ must satisfy detailed balance condition and ergodicity. 

In the following, we first discuss the update scheme for general interaction and then for
density-density iteration. The main difference between the two is the way to calculate local trace.

\subsection{General interactions}

As shown in Eq.~(\ref{wloc_ctx}), the calculation of local trace requires multiplication of matrices and is time-consuming.
We can take advantages of symmetries of $H_{\text{loc}}$ and divide the full Hilbert space of $H_{\text{loc}}$
into much smaller subspaces labeled by some good quantum numbers (GQNs)\cite{Haule_ctqmc_2007},
such as the total particle number $N_{\text{tot}}$, the total Spin $z$-component
$S_{\text{tot}}^{z}$, the total angular momentum $J_{z}$, etc. Single particle creation and annihilation
operators are therefore in block diagonal form, which speeds up the calculation. Further
speed-up can be achieved by using the divide-and-conquer\cite{iqist} trick based on
the fact that diagrammatic configurations are modified locally in
each update. 

\subsubsection{Pure-pair insertion/removal}

To insert a pure-pair in configuration $\mathcal{C}_{k}$, we pick
a random flavor $\alpha$, a random pair with type $s$, and a random time
$\tau$ in $(0,\beta)$. In the corresponding removal process, we
simply delete one of the existing pure-pairs among $k_{\alpha}+1$ pairs.
The ratio of the transition probabilities can be calculated for the inserting case as
\begin{equation}
\frac{p(k_{\alpha}\rightarrow k_{\alpha}+1)}{p(k_{\alpha}+1\rightarrow k_{\alpha})}=\frac{w_{\text{loc}}(\mathcal{C}_{k+1})\det(\mathcal{M}_{\alpha}^{(\mathcal{C}_{k+1})})^{-1}}{w_{\text{loc}}(\mathcal{C}_{k})\det(\mathcal{M}_{\alpha}^{(\mathcal{C}_{k})})^{-1}}\times\frac{2\beta}{k_{\alpha}+1},
\end{equation}
where $w_{\text{loc}}(\mathcal{C}_{k+1})$ is the local trace and $(\mathcal{M}_{\alpha}^{(\mathcal{C}_{k+1})})^{-1}$
is the hybridization matrix of the new configuration at order $k+1$.

\subsubsection{Left/right-exchange}

In the left-exchange update, we randomly pick a pair operator $s_{i}f_{\alpha_{i}}f_{\alpha_{i}^{\prime}}^{\dagger}$
together with its left neighbor $s_{i+1}f_{\alpha_{i+1}}f_{\alpha_{i+1}^{\prime}}^{\dagger}$
and exchange their annihilation operators if $\alpha_{i}\ne\alpha_{i+1}$
\begin{equation}
\begin{aligned}\cdots-s_{i+1}f_{\alpha_{i+1}}f_{\alpha_{i+1}^{\prime}}^{\dagger}(\tau_{i+1}) & -s_{i}f_{\alpha_{i}}f_{\alpha_{i}^{\prime}}^{\dagger}(\tau_{i})-\cdots\\
 & \Downarrow\\
\cdots-s_{i+1}f_{\alpha_{i}}f_{\alpha_{i+1}^{\prime}}^{\dagger}(\tau_{i+1}) & s_{i}f_{\alpha_{i+1}}f_{\alpha_{i}^{\prime}}^{\dagger}(\tau_{i})-\cdots.
\end{aligned}
\end{equation}
If $i=k$, the right-most pair is selected as the left neighbor of
$k$-th pair$ (k\ge2)$. It is equivalent to two successive shifts:
$f_{\alpha}$ from $\tau_{i+1}$ to $\tau_{i}$ and $f_{\alpha_{i+1}}$
from $\tau_{i}$ to $\tau_{i+1}$. Using Metropolis-Hasting algorithm
we obtain 
\begin{equation}
\frac{p(k)^{\prime}}{p(k)}=\frac{w_{\text{loc}}(\mathcal{C}_{k}^{\prime})}{w_{\text{loc}}(\mathcal{C}_{k})}\times\frac{\det(\mathcal{M}_{\alpha_{i}}^{(\mathcal{C}^{\prime}_{k})})^{-1}}{\det(\mathcal{M}_{\alpha_{i}}^{(\mathcal{C}_{k})})^{-1}}\times\frac{\det(\mathcal{M}_{\alpha_{i+1}}^{(\mathcal{C}^{\prime}_{k})})^{-1}}{\det(\mathcal{M}_{\alpha_{i+1}}^{(\mathcal{C}_k)})^{-1}},\label{lexchgDBC}
\end{equation}
 where $(\mathcal{M}_{\alpha}^{(\mathcal{C}^{\prime}_{k})})^{-1}$ ($\alpha=\alpha_{i},\alpha_{i+1}$)
is the new hybridization matrix of flavor $\alpha$ with shifted $f_{\alpha}$
 compared with original $(\mathcal{M}_{\alpha}^{(\mathcal{C}_{k})})^{-1}$.

The right-exchange updates works quite similar to left-exchange except that
it operates on creation operators, and the detailed balance condition
is of the form of Eq.~(\ref{lexchgDBC}) where $(\mathcal{M}_{\alpha}^{(\mathcal{C}^{\prime}_{k})})^{-1}$
is hybridization matrix with $f_{\alpha}^{\dagger}$ being shifted.

\subsubsection{Swap }

The $i$th pair is randomly selected, and we flip its type from $s_{i}$
to $-s_{i}$. Swap update is very important for satisfying ergodicity since it
switches virtual charge fluctuations between $f^{n}\rightarrow f^{n-1}$
and $f^{n}\rightarrow f^{n+1}$. Pure-pair will not be selected since the swap
of pure-pair can be done by removal of $s_{i}f_{\alpha_{i}}f_{\alpha_{i}^{\prime}}^{\dagger}$
and insertion of $-s_{i}f_{\alpha_{i}}f_{\alpha_{i}^{\prime}}^{\dagger}$
at $\tau_{i}$ successively. The ratio of the transition probabilities
is 
\begin{equation}
\frac{p(k)^{\prime}}{p(k)}=\frac{w_{\text{loc}}(\mathcal{C}_{k}^{\prime})}{w_{\text{loc}}(\mathcal{C}_{k})},
\label{swapmc}
\end{equation}
The reason why $\mathcal{M}^{-1}$ is not involved in Eq.~(\ref{swapmc}) is that it's block diagonal in 
spin-orbitals.

\subsection{Density-Density interactions}

If $H_{loc}$ commutes with the occupation number operator of each
orbital, the eigenstates of $H_{loc}$ are Fock states.
For each orbital, creation operator has to be followed by an annihilation operator 
for all valid configurations(we refer it as
NN-Rule). The weighting factor of the allowed configuration $\mathcal{C}_{k}$ can then be expressed as
\begin{equation}
\begin{aligned}w_{\text{loc}}(\mathcal{C}_{k}) & =s_{T_{\tau}}\cdot e^{-(\beta-\tau_{k})E_{\Gamma_{1}^{l}}}X_{\Gamma_{1}^{l}\Gamma_{k}^{l}}^{s_{k}\alpha_{k}\alpha_{k}^{\prime}}e^{-(\tau_{k}-\tau_{k-1})E_{\Gamma_{k}^{l}}}\cdots\\
 & \times X_{\Gamma_{3}^{l}\Gamma_{2}^{l}}^{s_{2}\alpha_{2}\alpha_{2}^{\prime}}e^{-(\tau_{2}-\tau_{1})E_{\Gamma_{2}^{l}}}X_{\Gamma_{2}^{l}\Gamma_{1}^{l}}^{s_{1}\alpha_{1}\alpha_{1}^{\prime}}e^{-(\tau_{1}-0)E_{\Gamma_{1}^{l}}}.
\end{aligned}
\end{equation}
To propose valid configurations, updates should be carefully designed
to satisfy the NN-Rule. 

\subsubsection{Pure-pair insertion/removal}

The main difference with the general interaction case is that the pair type $s$
can not be randomly selected. For a given configuration, if the orbital $\alpha$ is occupied (unoccupied) in the Fock state
spanning $\tau$,  only the insertion of $f_{\alpha}^{\dagger}f_{\alpha}$($f_{\alpha}f_{\alpha}^{\dagger}$)
at $\tau$ is allowed. When it comes to
pure-pair removal, we correspondingly delete $f_{\alpha}^{\dagger}f_{\alpha}$ ($f_{\alpha}^{\dagger}f_{\alpha}$) away from $\tau$. The condition for detail balance reads 
\begin{equation}
\frac{p(k_{\alpha}\rightarrow k_{\alpha}+1)}{p(k_{\alpha}+1\rightarrow k_{\alpha})}=\frac{w_{\text{loc}}(\mathcal{C}_{k+1})\det(\mathcal{M}_{\alpha}^{(\mathcal{C}_{k+1})})^{-1}}{w_{\text{loc}}(\mathcal{C}_{k})\det(\mathcal{M}_{\alpha}^{(\mathcal{C}_{k})})^{-1}}\times\frac{\beta}{k_{\alpha}+1}.
\end{equation}

\subsubsection{Left/right-exchange and swap}

Exchange process which violate the NN-Rule will be directly rejected. Swap updates
will not violate the rule since only mix-pairs are swapped. The conditions for the detailed balance
 are same with those of general interactions except for the calculations
of local trace. While left/right-exchange is equivalent to
shift of segments, swap is equivalent to switch between infinitesimal  small
segment and anti-segment. 

\section{Measurements}

The most important observable for QMC impurity solvers is the finite temperature imaginary-time Green's function defined by
$G_{\alpha\alpha'}^{f}(\tau)=-\langle\mathcal{T}_{\tau}f_\alpha(\tau)f_{\alpha'}^{\dagger}(0)\rangle$. 
The single particle green's function, in general, includes high energy process that involve states with different occupation numbers. Such process, however, are missing in our approximated parition function, Eq.~(\ref{zapprox}), in the Kondo limit, where charge fluctuations are projected out completely. Nevertheless, we can still measure the low-energy contributions to $G_{\alpha\alpha'}^{f}(\tau)$, which correspond to the quasi-particle part in the single-particle excitations. 
Here, we give a brief descriptions of how to measure $G_{\alpha\alpha'}^{f}(\tau)$.
The step by step derivation of the measurement formula is given in Appendix~B.

\begin{figure}[htp]
\includegraphics[clip,width=3.4in,angle=0]{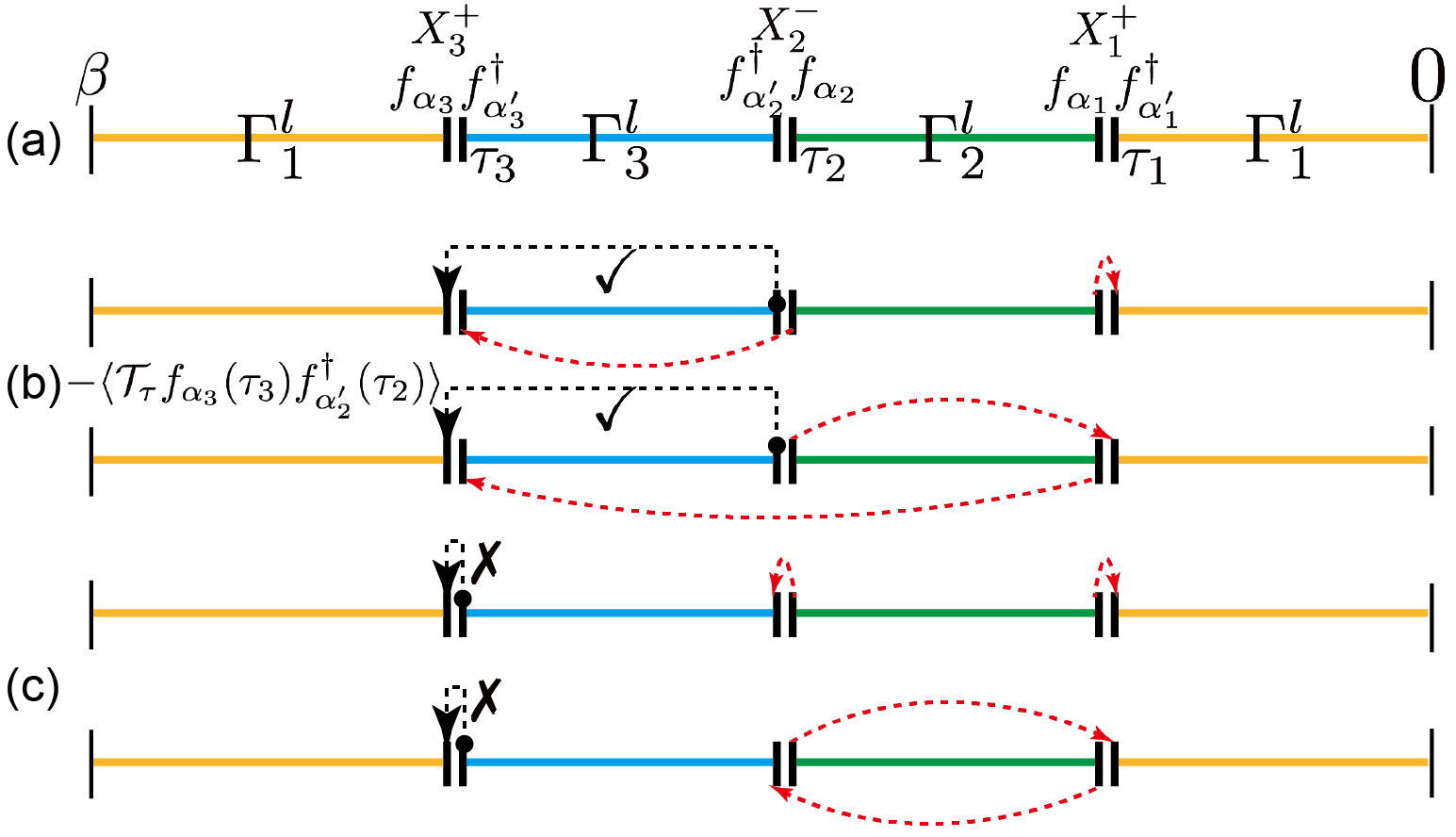}
\caption{(Color online) Schematic plot of a third order configuration for measurement of Green's function in Kondo regime.
(a) Local trace part of Green's function $G_{\alpha\alpha}^{f}(\tau)$, which is the same as that of $Z$ shown in Fig.~\ref{diagram_c3}(a).
(b) An example of allowed (marked by \checkmark) measurement for $G_{\alpha\alpha}^{f}(\tau)$. Two operators from two different $X$
matrices [here $f_{\alpha_3}(\tau_3)$ of pair $X_3$ and $f_{\alpha^\prime_2}^\dagger(\tau_2)$ of pair $X_2$] are identified
as operators in $G_{\alpha\alpha}^{f}(\tau)$. In upper and lower panels, we plot existing hybridization lines which do not connect between the selected operators.
(c) An example of forbidden (marked by \xmark) measurement in $G_{\alpha\alpha}^{f}(\tau)$. Two operators belonging to the same $X$ matrix are not allowed to be 
chosen in measurement.}
\label{Greenfun_c3}
\end{figure}

In the CTQMC, the diagrams contributing to $G_{\alpha\alpha'}^{f}(\tau)$ can be generated from diagrams in $Z$: One chooses an arbitrary pair of creation and annihilation operators in a given configuration $\mathcal{C}_{k}$, and removes the corresponding contributions to the determinant $\Delta_\alpha^{(\mathcal{C}_{k})}$.
Within the approximation applied to the partition function, only a specific pairs of creation and annihilation operators have contributions to $G_{\alpha\alpha'}^{f}(\tau)$. A pair of operators that belong to different $X$ matrices does contribute, while those on the same $X$ matrix do not. Those diagrams are illustrated in Fig.~\ref{Greenfun_c3}(b) and Fig.~\ref{Greenfun_c3}(c), respectively.

The measurement formula is thus given by
\begin{equation}
G^f_{\alpha\alpha^{\prime}}(\tau)=-\langle\frac{1}{\beta} \sideset{}{'}\sum_{m,n=1}^{k} \delta_{\alpha_{m},\alpha}\delta_{\alpha_{n}^{\prime},\alpha^{\prime}}\delta^{-}[\tau-(\tau_{m}-\tau_{n})]\mathcal{M}_{nm}^{(\mathcal{C}_{k})}\rangle_{Z},
\label{G_KR}
\end{equation}
where $\prime$ stands for the resctriction of summations to $m \ne n$ ( $m$ and $n$ are one different $X$ matrices). The function $\delta^-$ is defined by $\delta^{-}(\tau-\tau^\prime)=\delta(\tau-\tau^\prime)$ for $\tau>0$, and $\delta^{-}(\tau-\tau^\prime)=-\delta(\tau-\tau^\prime-\beta)$ for $\tau<0$. 
After the Fourier transform, we obtain
\begin{equation}
G_{\alpha\alpha^\prime}^{f}(i\omega_{l})=-\langle\frac{1}{\beta} \sideset{}{'}\sum_{m,n=1}^{k} \delta_{\alpha_{m},\alpha}\delta_{\alpha_{n}^{\prime},\alpha^{\prime}}\mathcal{M}_{nm}^{(\mathcal{C}_{k})}e^{i\omega_{l}(\tau_{m}-\tau_{n})}\rangle.
\end{equation}

We can compare the present measurement formula, Eq.~(\ref{G_KR}), with that for the of $t$-matrix in the CS model, Eq.~(9) in Ref.~[\onlinecite{ctJ_paper}]. 
They are related by $t(\tau)=V^2G(\tau)$ if there is no $k$ dependence in $V_k$.

The asymptotic behavior of $G(i\omega_n)$ is $i\omega_n * G(i\omega_n) |_{n\rightarrow\infty}=z$ with $z<1$ being the quasi-particle weight in the Kondo limit.

The measurement formula for the two particle correlation function bear exactly the same form as Eq.~(11)-(13) of Ref.~[\onlinecite{ctJ_paper}], which are not mentioned here.

\section{Benchmarks}

While Coqblin-Shrieffer(CS) model is a low energy effective Hamiltonian
of ASIM in large $U$ limit in which only virtual excitations $f^{1}\rightarrow f^{0}$
survive, Kondo model incorporates both $f^{1}\rightarrow f^{0}$
and $f^{1}\rightarrow f^{2}$ by assuming deep impurity level $\epsilon_{f}$
and large $U$. Both the two models can be derived
by SWT from SIAM with density-density interaction shown below
\begin{equation}
\begin{aligned}H & =\sum_{k\alpha}\epsilon_{k}c_{k\alpha}^{\dagger}c_{k\alpha}+\sum_{\alpha=-j}^{j}\varepsilon_{\alpha}n_{\alpha}+U\sum_{\alpha<\alpha^{\prime}}n_{\alpha}n_{\alpha^{\prime}},\\
 & +\sum_{\alpha k}[V_{k}^{\alpha}f_{\alpha}^{\dagger}c_{k\alpha}+V_{k}^{\alpha*}c_{k\alpha}^{\dagger}f_{\alpha}],
\end{aligned}
\end{equation}
with $N=2j+1$. A constant density of states $\rho(\epsilon)=\frac{1}{2D}\theta(D-|\epsilon|)$
with $D=1$ is chosen for conduction electrons. Both the CS and Kondo models are derived from SIAM
under certain conditions. Therefore the comparison between the  Monte Carlo simulations on these models using CT-J method and directly on SIAM using our new method proposed in this paper can be used as the benchmark.
For sake of simplicity, our CT-QMC formalism for partition function (\ref{zapprox})
is referred as CT-X, where ``X'' refers to $X$-matrices.

\subsection{CS model}

The CS model reads 
\begin{align}
H_{0}&=\sum_{k\alpha}\epsilon_{k}c_{k\alpha}^{\dagger}c_{k\alpha}+\sum_{\alpha}(\varepsilon_{\alpha}+J_{\alpha\alpha})|\alpha\rangle\langle\alpha|,
\\
H_{1}&=\sum_{\alpha\alpha^{\prime}}J_{\alpha\alpha^{\prime}}|\alpha\rangle\langle\alpha^{\prime}|(-c_{\alpha}c_{\alpha^{\prime}}^{\dagger}),
\end{align}
where $c_{\alpha}=N_{0}^{-1/2}\sum_{k}c_{k\alpha}$, with $N_{0}$
being number of sites and $\alpha$ denotes the spin/orbital indices. 
Partition function of CS model
in CT-J can be obtained by applying the following restrictions to
Eq.~(\ref{xmatm1}) and Eqs.~(\ref{zapprox})--(\ref{wloc_ctx}):
\begin{itemize}
\item $V_{k}^{\alpha}=V_{k}^{\alpha*}=V^{\alpha}$, since exchange parameters
in CS model can be chosen as momentum independent;
\item $J_{\alpha\alpha^{\prime}}\equiv\frac{V_{\alpha}V_{\alpha^{\prime}}}{-\min\{\varepsilon_{\alpha}\}}$,
where $E_{\Gamma^{h}}=-\min\{\varepsilon_{\alpha}\}$ is the shifted
energy of $f^{0}$ state;
\item $X^{-1,\alpha\alpha^{\prime}}$ has only one none-zero element $X_{|\alpha\rangle|\alpha^{\prime}\rangle}^{-1,\alpha\alpha^{\prime}}=\frac{1}{-\min\{\varepsilon_{\alpha}\}}|\alpha\rangle\langle\alpha^{\prime}|$.
\end{itemize}
To simulate the CS model, we should put an additional restriction
to $\mathcal{C}_{k}$ with $\{s_{i}=-1|i=1,\cdots,k\}$, which means
that only pair operators describing $f^{1}\rightarrow f^{0}$ enters
in $\mathcal{C}_{k}$. Furthermore, intra-pair swap update is forbidden
since it gives rise to virtual excitation $f^{1}\rightarrow f^{2}$
which is absent in the CS model. 

\subsubsection{t-matrix}

\begin{figure}[htp]
\includegraphics[clip,width=3.4in,angle=0]{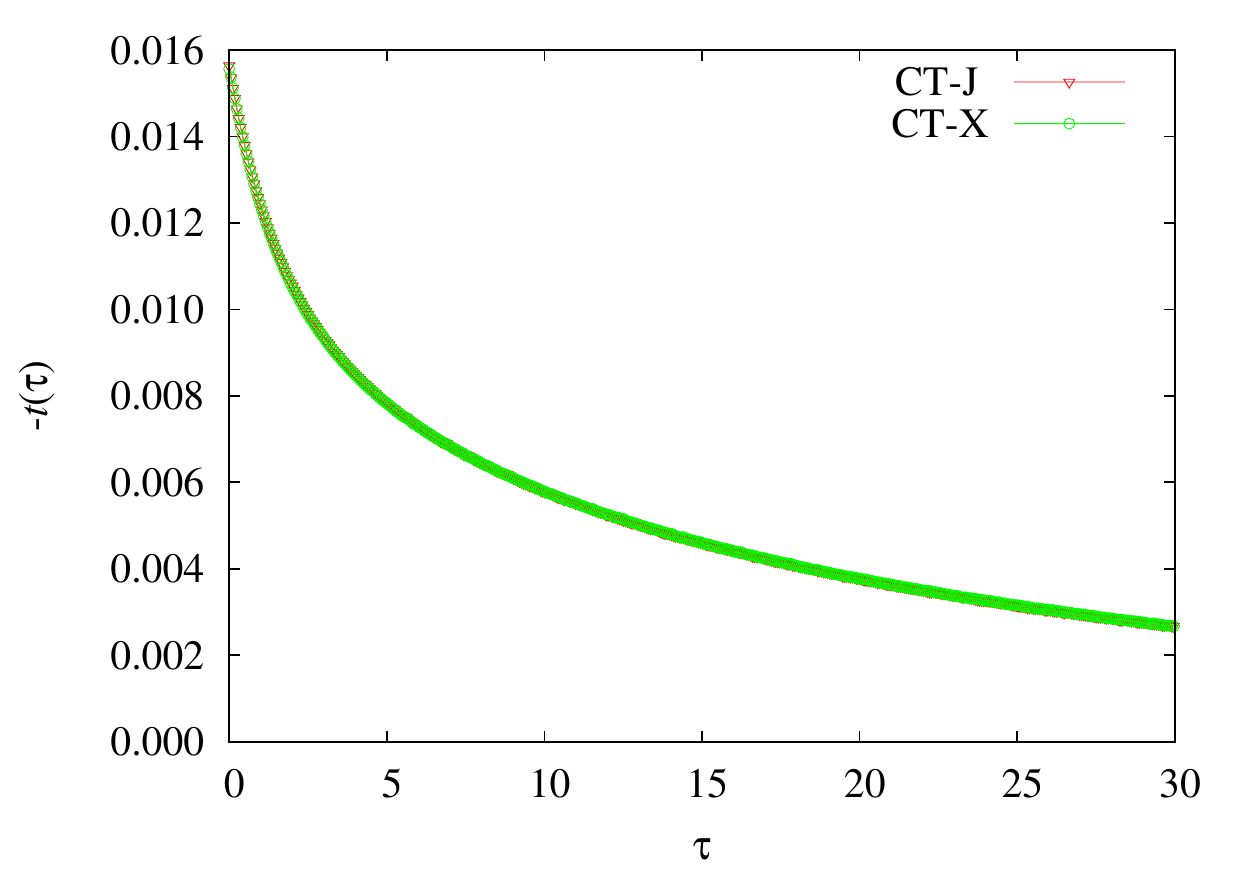}
\caption{The impurity t-matrix $t_\alpha(\tau)$ in the imaginary-time domain for $N=8$, $J=0.075$ and $T=0.001$. Datas of CT-J is 
obtained from Fig.~7 in Ref.~[\onlinecite{ctJ_paper}].} \label{fig1}
\end{figure}

To test CT-X, we calculate t-matrix with $N=8$ and compare it with the results obtained by CT-J. We choose
the exchange parameter $J=\frac{V^{2}}{-\epsilon_{f}}$ to be 0.075 and temperature $T=0.001D$.
The results obtained by CT-X and CT-J are plotted together in FIG.~\ref{fig1}, which show excellent agreement.

\subsubsection{static susceptibility }

\begin{figure}[htp]
\includegraphics[clip,width=3.4in,angle=0]{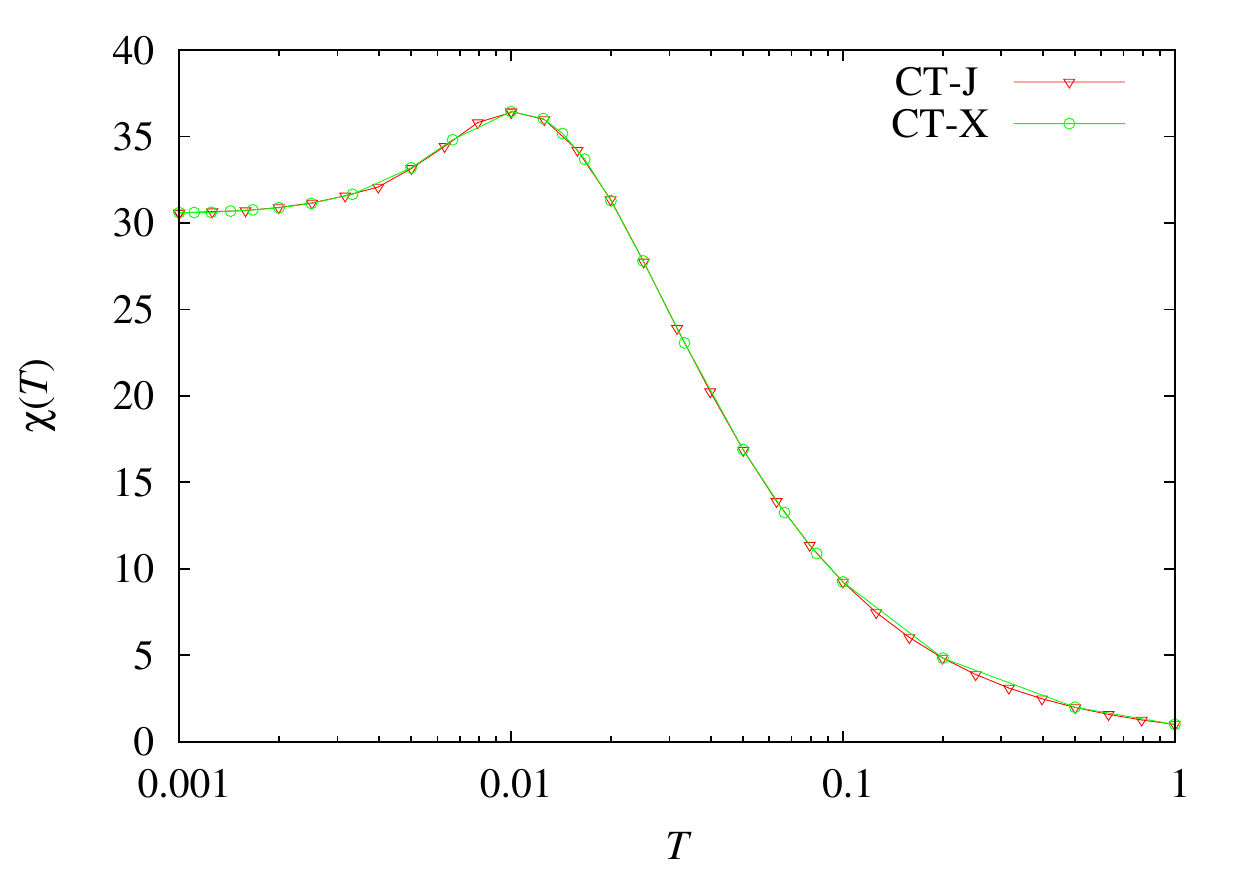}
\caption{Temperature dependence of the static susceptibility $\chi$ for N=8 and J=0.075. Datas of CT-J is 
obtained from Fig.~6 in Ref.~[\onlinecite{ctJ_paper}].} \label{fig2}
\end{figure}

The static susceptibility is evaluated by integrating dynamical susceptibility
$\chi(\tau)$ as introduced in detail in section 2.3 of Ref.~[\onlinecite{ctJ_paper}]. 
The results obtained by CT-X and CT-J are shown in FIG.~\ref{fig2}. Again they match each other very well.

\subsection{Kondo model}

The Kondo model is given by 
\begin{equation}
H=\sum_{k\sigma}\epsilon_{k}c_{k\sigma}^{\dagger}c_{k\sigma}+J\boldsymbol{S}\cdot\boldsymbol{\sigma}_{c}.
\end{equation}
where $\boldsymbol{S}=\sum_{\alpha\beta}f_{\alpha}^{\dagger}\boldsymbol{\sigma}_{\alpha\beta}f_{\beta}$
and $\boldsymbol{\sigma}_{c}=\sum_{\sigma\sigma^{\prime}}c_{\sigma}^{\dagger}\boldsymbol{\sigma}_{\sigma\sigma^{\prime}}c_{\sigma^{\prime}}$ denoting the spin operators of the local moments and itinerant electrons
respectively.
The spin-spin exchange terms can be obtained by considering both of the two virtual processes
 $f^{1}\rightarrow f^{0}$ and $f^{1}\rightarrow f^{2}$.
With the particle-hole symmetry, we set $U=-2\epsilon_{f}$ thus $J=\frac{V^{2}}{-\epsilon_{f}}=\frac{V^{2}}{U+\epsilon_{f}}$
. To simulate the Kondo model by our CT-X method, all types of pair operators are allowed
to appear in the MC configurations. 

\begin{figure}[htp]
\includegraphics[clip,width=3.4in,angle=0]{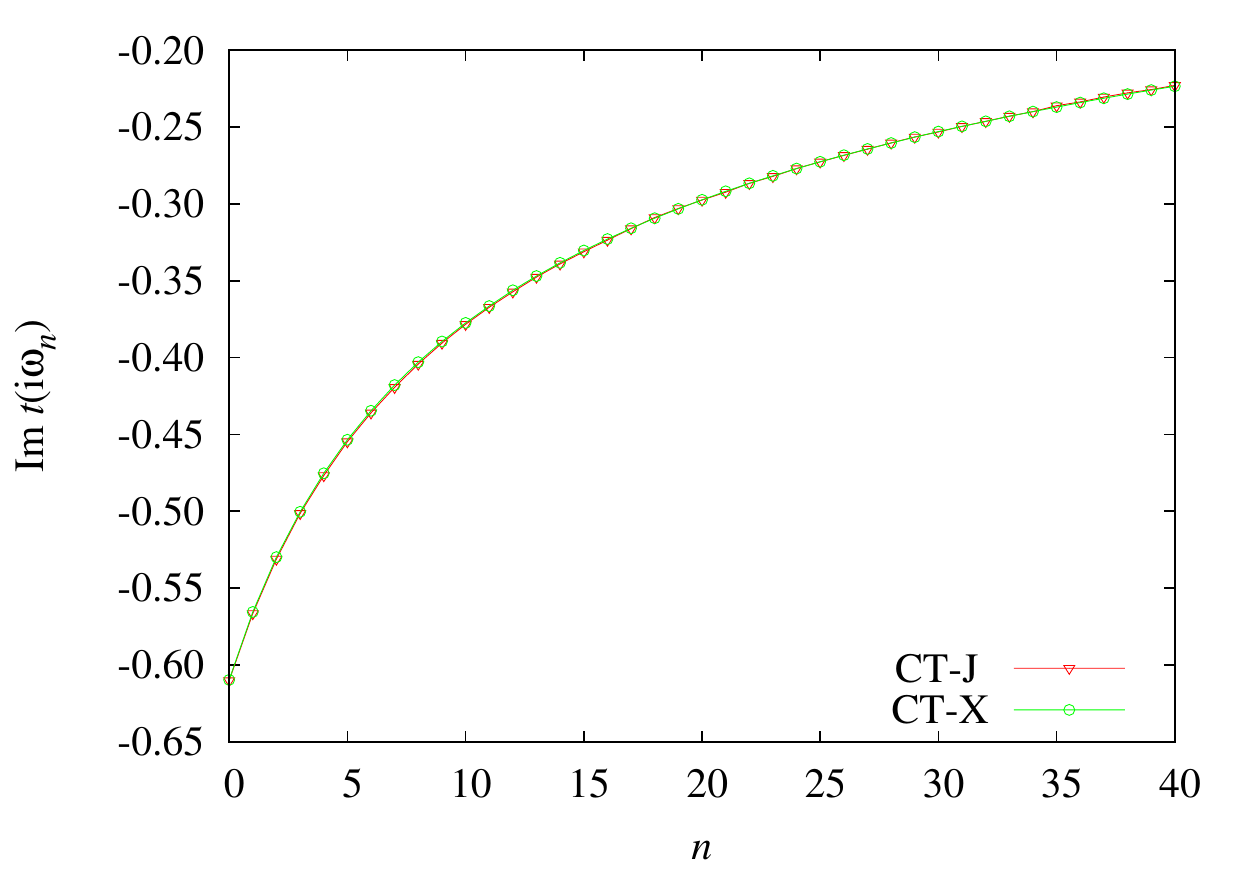}
\caption{Imaginary part of the impurity t-matrix $t(i\omega_n)$ of the Kondo model in the imaginary-frequency domain for N=2 and J=0.300. 
Note: datas of CT-J in this figure is collected by means of WebPlotDigitizer~\cite{webpicdigit} from Fig.~11. in Ref.~[\onlinecite{ctJ_paper}]. } \label{fig3}
\end{figure}

As a benchmark, we calculated the $t$-matrix with $J$=0.3 and $T$=0.001 by CX-T and compare it with the results obtained by CT-J In FIG.~\ref{fig3}.  Again the results from CT-X and CT-J agree very well indicating that CT-X can 
treat two types of virtual charge fluctuations well. With the particle-hole symmetry of Kondo model, the real part of $t(i\omega_n)$ is zero and hence not plotted in FIG.~\ref{fig3}. 

\subsection{Kondo lattice model (KLM)}
\begin{figure}[htp]
\includegraphics[clip,width=3.4in,angle=0]{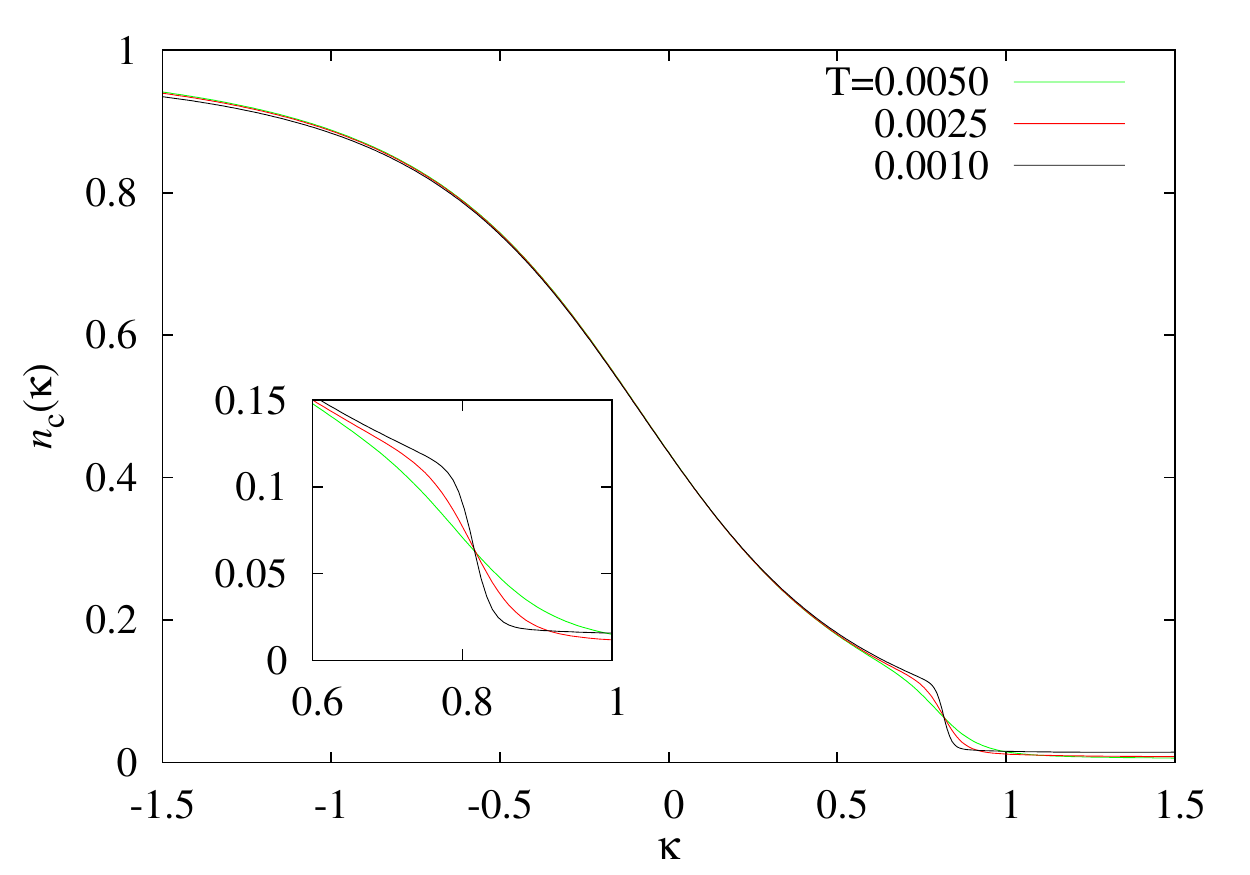}
\caption{(Color online). Temperature dependence of momentum distribution $n_{c}(\boldsymbol{\kappa})$ computed by CT-X for $J$=0.3 and $n_c$=0.9. The vicinity of the large Fermi surface is enlarged in the inset.}\label{fig4}
\end{figure}

\begin{figure}[htp]
\includegraphics[clip,width=3.4in,angle=0]{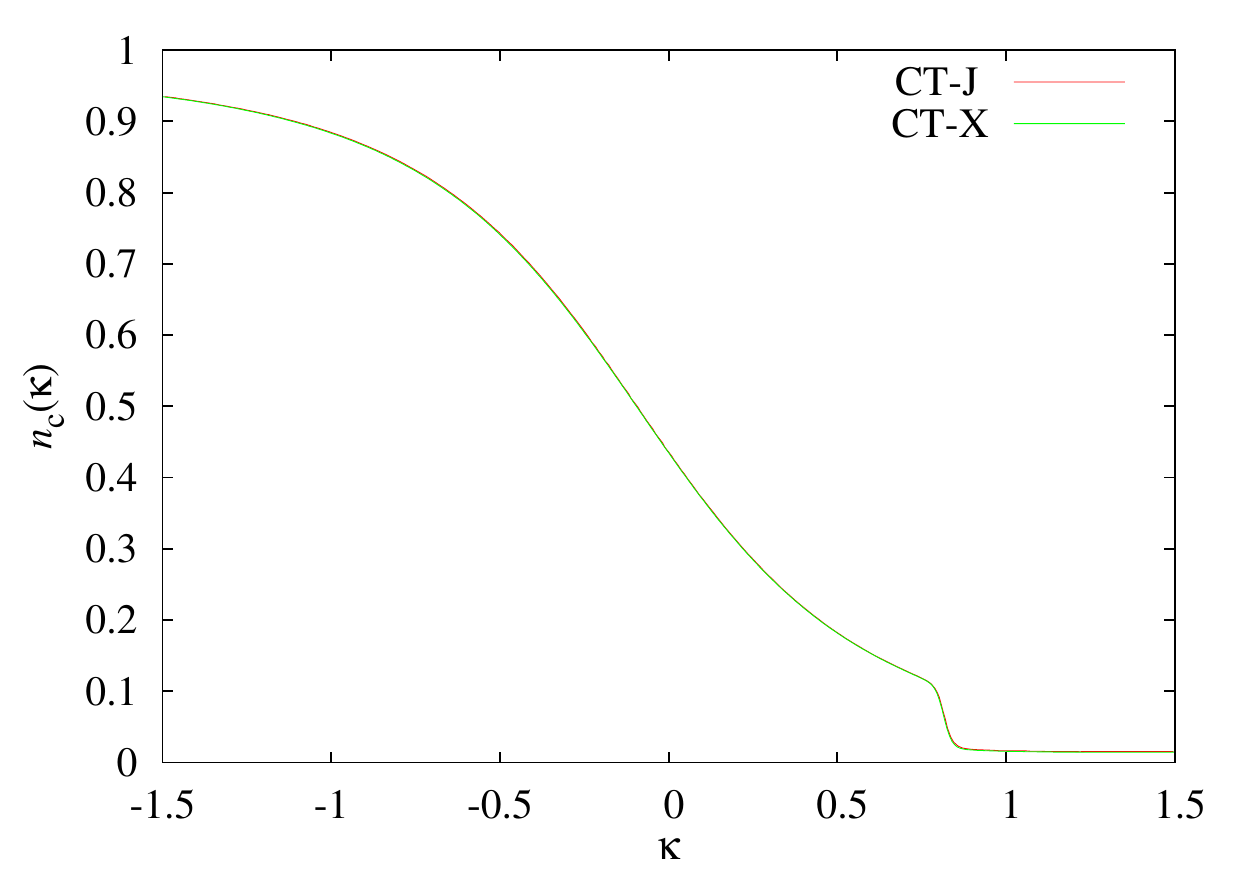}
\caption{(Color online). Momentum distribution $n_{c}(\boldsymbol{\kappa})$ at T=0.001 and comparison with the CT-X. Datas of CT-J is 
obtained from Fig.~74 in Ref.~[\onlinecite{otsuki_prl2009}]. }\label{fig5}
\end{figure}

KLM reads 
\begin{equation}
H=\sum_{k\sigma}\epsilon_{k}c_{k\alpha}^{\dagger}c_{k\alpha}+J\sum_{i}S_{i}\cdot\boldsymbol{\sigma}_{i}.
\end{equation}
To further test this new impurity solver, we perform DMFT calculations
on KLM in the infinite-dimension hyper-cubic lattice
with the density of states $\rho_{c}(\omega)=D^{-1}\sqrt{2/\pi}exp(-2\omega^{2}/D^{2})$.
We set $D=1$ and fix the conduction-electron density per site as
$n_{c}=0.9$ as that in Ref.~[\onlinecite{otsuki_prl2009}]. The DMFT is iterated
on conduction-electron self-energy $\Sigma_{c}$$(i\omega_{n})$,
which is related to the cavity Green function $\mathcal{G}_{c}^{0}$
and the measured impurity $t$-matrix by Dyson equation $\Sigma_{c}(i\omega_{n})^{-1}=t(i\omega_{n})+\mathcal{G}_{c}^{0}(i\omega_{n})$.
For sake of benchmark, we calculate the momentum distribution of conduction electrons: 
\begin{equation}
n_{c}(\boldsymbol{\kappa})=\langle c_{k\sigma}^{\dagger}c_{k\sigma}\rangle=T\sum_{n}G_{c}(\boldsymbol{\kappa},i\omega_{n})e^{i\omega_{n}0^{+}},
\end{equation}
where $\boldsymbol{\kappa}\equiv\epsilon_{k}$ and $G_{c}$ is the
conduction-electron Green function in the KLM, $G_{c}(\boldsymbol{\kappa},i\omega_{n})=[i\omega_{n}-\boldsymbol{\kappa}+\mu-\Sigma_{c}(i\omega_{n})]^{-1}$. 
FIG.~\ref{fig4} shows the temperature dependence of $n_{c}(\boldsymbol{\kappa})$ at T=0.0050, 0.0025 and 0.0010 and well 
reproduces the evolution of Fermi surface as shown in FIG. 4 of Ref.~[\onlinecite{otsuki_prl2009}]. For comparison, we plot in FIG.~\ref{fig5} the results computed by CT-X together with results
computed by CT-X at T=0.001. Once again it demonstrates that CT-X can treat two types of virtual charge fluctuations well. 

\section{Discussion and Conclusion}

We have proposed a new CTQMC method called CT-X, which can simulate the SIAM in the Kondo limit
 by projecting out local charge fluctuations, not in the effective Hamiltonian but each diagram sampled by the MC procedure. 
This is done by approximating the high-energy states' imaginary-time
evolution operators which are sharply decreasing by a probability normalized $\delta$ function.
This approximation is equivalent to apply SWT for each particular diagrams.

Benchmarks of CT-X on CS model, Kondo model and Kondo lattice model with previously proposed
CT-J method show that CT-X method works very well for these model systems. 
However, since in the CT-X method the SWT type approximation is applied to each particular 
Feynman diagrams in Monte Carlo procedure, it can be easily applied to more general quantum impurity models 
that describe realistic materials. Realistic models contain a generalized form of interaction, generalized
occupation number and generalized crystal field, which is difficult for the method based on effective model approach
such as the CT-J method. Therefore the CT-X method developed in the present paper can become a very good 
impurity solver for DMFT to study strongly correlated systems such as the heavy fermion materials.

\newpage
\begin{widetext}
\appendix
\numberwithin{equation}{section}
\numberwithin{figure}{section}
\newpage
\section{Partition function}
In this appendix, we derive the starting expression for $Z$ in Eq.~(\ref{Z_original}).
First, $Z=\text{Tr}[e^{-\beta H}]$ is perturbatively expanded in terms of $H_1$ as
\begin{equation}
Z=\text{Tr}[e^{-\beta H_{0}}\mathcal{T}_{\tau}e^{-\int_{0}^{\beta}H_{1}(\tau)d\tau}]=\sum_{n=0}^{\infty}(-1)^{n}\frac{1}{n!}\int_{0}^{\beta}d\tau_{1}\cdots\int_{0}^{\beta}d\tau_{n}\text{Tr}[\mathcal{T}_{\tau}e^{-\beta H_{0}}H_{1}(\tau_{1})\cdots H_{1}(\tau_{k})],
\end{equation}
where $H_{0}=H_{\text{loc}}+H_{\text{bath}}$, $H_{1}\equiv H_{\text{hyb}}=V+V^{\dagger}$
with $V\equiv\sum_{p}V_{p}^{\alpha}c_{p}^{\dagger}f_{\alpha}$ and
$p\equiv\boldsymbol{k}\nu$. Particle number conservation requires
that the terms with the non-zero contribution to $Z$ must contain
an equal number of $V$ and $V^{\dagger}$. In other words, $n$ needs
to be even. By denoting $n=2k$, we have 
\begin{equation}
Z=\sum_{k=0}^{\infty}\frac{1}{(2k)!}\int_{0}^{\beta}d\tau_{1}\cdots\int_{0}^{\beta}d\tau_{2k}\text{Tr}[\mathcal{T}_{\tau}e^{-\beta H_{0}}H_{1}(\tau_{1})\cdots H_{1}(\tau_{2k})].
\end{equation}
There are $C_{2k}^k$ different ways to divide $2k$ $H_1$ terms into two groups and pick $V$ part from the first group and $V^\dagger$ part from the other. 
To label the $i$-th configuration, we
introduce an integer set with $k$ elements, $S^{i}$, to mark the group of $H_1$ terms from which the $V$ terms have been picked. 
These integers ranging from 1 to $2k$ are pairwise distinct
and arranged in ascending order. If we shift number of $k$ $V$ to
the left side, there will be no additional sign created since $V$ and $V^{\dagger}$
are bosonic operators
\begin{equation}
Z=\sum_{k=0}^{\infty}\frac{1}{(2k)!}\int_{0}^{\beta}d\tau_{1}\cdots\int_{0}^{\beta}d\tau_{2k}\sum_{i=1}^{C_{2k}^{k}}\text{Tr}[\mathcal{T}_{\tau}e^{-\beta H_{0}}\prod_{j=1}^{k}V(\tau_{S_{j}^{i}})\prod_{j=1}^{k}V^{\dagger}(\tau_{\overline{S^{i}}_{j}})],
\end{equation}
 where $\overline{S^{i}}$ is used to denote the complement
set of $S^{i}$ in set $\{1,2,\cdots,2k\}$. All these $C_{2k}^{k}$ terms
contribute equally to $Z$ because of unconstrained integrals and
bosonic feature of $V$ and $V^{\dagger}$. Then we have
\begin{equation}
\begin{aligned}
Z & =\sum_{k=0}^{\infty}\frac{C_{2k}^{k}}{(2k)!}\int_{0}^{\beta}d\tau_{1}\cdots\int_{0}^{\beta}d\tau_{2k}\text{Tr}[\mathcal{T}_{\tau}e^{-\beta H_{0}}V(\tau_{1})\cdots V(\tau_{k})V^{\dagger}(\tau_{k+1})\cdots V^{\dagger}(\tau_{2k})]\\
 & =\sum_{k=0}^{\infty}\frac{1}{(k!)^{2}}\int_{0}^{\beta}d\tau_{1}\cdots\int_{0}^{\beta}d\tau_{k}\int_{0}^{\beta}d\tau_{1}^{\prime}\cdots\int_{0}^{\beta}d\tau_{k}^{\prime}\text{Tr}[\mathcal{T}_{\tau}e^{-\beta H_{0}}V(\tau_{1})\cdots V(\tau_{k})V^{\dagger}(\tau_{1}^{\prime})\cdots V^{\dagger}(\tau_{k}^{\prime})]\\
 & =\sum_{k=0}^{\infty}\int_{0}^{\beta}d\tau_{1}\cdots\int_{\tau_{k-1}}^{\beta}d\tau_{k}\int_{0}^{\beta}d\tau_{1}^{\prime}\cdots\int_{\tau_{k-1}^{\prime}}^{\beta}d\tau_{k}^{\prime}\text{Tr}[\mathcal{T}_{\tau}e^{-\beta H_{0}}V(\tau_{1})\cdots V(\tau_{k})V^{\dagger}(\tau_{1}^{\prime})\cdots V^{\dagger}(\tau_{k}^{\prime})]
\end{aligned}
\end{equation}
where $\{\tau_{k+1},\cdots,\tau_{2k}\}$ is renamed as $\{\tau_{1}^{\prime},\cdots,\tau_{k}^{\prime}\}$
in the second step, while the unconstrained integrals are replaced
by the constrained ones in the third step. Plugging the explicit forms of $V$ and $V^{\dagger}$ into 
$Z$, we obtain 
\begin{equation}
\begin{aligned}
Z & =\sum_{k=0}^{\infty}\int_{0}^{\beta}d\tau_{1}\cdots\int_{\tau_{k-1}}^{\beta}d\tau_{k}\int_{0}^{\beta}d\tau_{1}^{\prime}\cdots\int_{\tau_{k-1}^{\prime}}^{\beta}d\tau_{k}^{\prime}\sum_{\substack{\alpha_{1}\cdots\alpha_{k}\\
\alpha_{1}^{\prime}\cdots\alpha_{k}^{\prime}
}
}\sum_{\substack{p_{1}\cdots p_{k}\\
p_{1}^{\prime}\cdots p_{k}^{\prime}
}
}V_{p_{1}}^{\alpha_{1}}\cdots V_{p_{k}}^{\alpha_{k}}V_{p_{1}^{\prime}}^{\alpha_{1}^{\prime}}\cdots V_{p_{k}^{\prime}}^{\alpha_{k}^{\prime}}\\
 & \times\text{Tr}[\mathcal{T}_{\tau}e^{-\beta H_{0}}c_{p_{1}}^{\dagger}(\tau_{1})f_{\alpha_{1}}(\tau_{1})\cdots c_{p_{k}}^{\dagger}(\tau_{k})f_{\alpha_{k}}(\tau_{k})f_{\alpha_{1}^{\prime}}^{\dagger}(\tau_{1}^{\prime})c_{p_{1}^{\prime}}(\tau_{1}^{\prime})\cdots f_{\alpha_{k}^{\prime}}^{\dagger}(\tau_{k}^{\prime})c_{p_{k}^{\prime}}(\tau_{k}^{\prime})].
\end{aligned}
\end{equation}
If we move all the conduction electrons' operators in $\text{Tr}[\mathcal{T}_{\tau}e^{-\beta H_{0}}\cdots]$
to the left side without altering their numerical orders, an extra
sign $s$ relating to such manipulation will arise since we are dealing
with fermionic operators. Fortunately, $s$ turns out to be 1 
\begin{equation}
s=(-1)^{k}\cdot(-1)^{\sum_{i=1}^{2k}(i-1)}=(-1)^{2k^{2}}=1,
\end{equation}
resulting in
\begin{equation}
\begin{aligned}
 & \text{Tr}[\mathcal{T}_{\tau}e^{-\beta H_{0}}c_{p_{1}}^{\dagger}(\tau_{1})f_{\alpha_{1}}(\tau_{1})\cdots c_{p_{k}}^{\dagger}(\tau_{k})f_{\alpha_{k}}(\tau_{k})f_{\alpha_{1}^{\prime}}^{\dagger}(\tau_{1}^{\prime})c_{p_{1}^{\prime}}(\tau_{1}^{\prime})\cdots f_{\alpha_{k}^{\prime}}^{\dagger}(\tau_{k}^{\prime})c_{p_{k}^{\prime}}(\tau_{k}^{\prime})]\\
= & \text{Tr}[\mathcal{T}_{\tau}e^{-\beta H_{0}}c_{p_{1}}^{\dagger}(\tau_{1})\cdots c_{p_{k}}^{\dagger}(\tau_{k})c_{p_{1}^{\prime}}(\tau_{1}^{\prime})\cdots c_{p_{k}^{\prime}}(\tau_{k}^{\prime})f_{\alpha_{1}}(\tau_{1})\cdots f_{\alpha_{k}}(\tau_{k})f_{\alpha_{1}^{\prime}}^{\dagger}(\tau_{1}^{\prime})\cdots f_{\alpha_{k}^{\prime}}^{\dagger}(\tau_{k}^{\prime})].
\end{aligned}
\end{equation}
Separating the bath and impurity operators, we obtain
\begin{equation}
\begin{aligned}
Z & =\sum_{k=0}^{\infty}\int_{0}^{\beta}d\tau_{1}\cdots\int_{\tau_{k-1}}^{\beta}d\tau_{k}\int_{0}^{\beta}d\tau_{1}^{\prime}\cdots\int_{\tau_{k-1}^{\prime}}^{\beta}d\tau_{k}^{\prime}\sum_{\substack{\alpha_{1}\cdots\alpha_{k}\\
\alpha_{1}^{\prime}\cdots\alpha_{k}^{\prime}
}
}\sum_{\substack{p_{1}\cdots p_{k}\\
p_{1}^{\prime}\cdots p_{k}^{\prime}
}
}V_{p_{1}}^{\alpha_{1}}\cdots V_{p_{k}}^{\alpha_{k}}V_{p_{1}^{\prime}}^{\alpha_{1}^{\prime}}\cdots V_{p_{k}^{\prime}}^{\alpha_{k}^{\prime}}\\
 & \times\text{Tr}_{c}[\mathcal{T}_{\tau}e^{-\beta H_{\text{bath}}}c_{p_{1}}^{\dagger}(\tau_{1})\cdots c_{p_{k}}^{\dagger}(\tau_{k})c_{p_{1}^{\prime}}(\tau_{1}^{\prime})\cdots c_{p_{k}^{\prime}}(\tau_{k}^{\prime})]\\
 & \times\text{Tr}_{f}[\mathcal{T}_{\tau}e^{-\beta H_{\text{loc}}}f_{\alpha_{1}}(\tau_{1})\cdots f_{\alpha_{k}}(\tau_{k})f_{\alpha_{1}^{\prime}}^{\dagger}(\tau_{1}^{\prime})\cdots f_{\alpha_{k}^{\prime}}^{\dagger}(\tau_{k}^{\prime})].
\end{aligned}
\label{Ztmp}
\end{equation}
According to Eq.~(\ref{Ztmp}), configuration space of $Z$ is given by sets of
imaginary times and corresponding orbitals
\begin{equation}
\mathcal{C}=\{\{\tau_{1},\cdots,\tau_{k},\tau_{1}^{\prime},\cdots,\tau_{k}^{\prime}\},\{\alpha_{1},\cdots,\alpha_{k},\alpha_{1}^{\prime},\cdots,\alpha_{k}^{\prime}\}|k=0,1,\cdots\},
\end{equation}
 with $\tau_{1}<\cdots<\tau_{k}$ and $\tau_{1}^{\prime}<\cdots<\tau_{k}^{\prime}$.
Furthermore, we introduce the definition of hybridization determinant
for configuration $\mathcal{C}_{k}$ as
\begin{equation}
\begin{aligned}
\det\Delta^{(\mathcal{C}_{k})} & \equiv\det\left(\begin{array}{ccc}
\triangle_{\alpha_{1}^{\prime}\alpha_{1}}(\tau_{1}^{\prime}-\tau_{1}) & \cdots & \triangle_{\alpha_{1}^{\prime}\alpha_{k}}(\tau_{1}^{\prime}-\tau_{k})\\
\vdots & \ddots & \vdots\\
\triangle_{\alpha_{k}^{\prime}\alpha_{1}}(\tau_{k}^{\prime}-\tau_{1}) & \cdots & \triangle_{\alpha_{k}^{\prime}\alpha_{k}}(\tau_{k}^{\prime}-\tau_{k})
\end{array}\right)\\
 & =\frac{1}{Z_{\text{bath}}} \sum_{\substack{p_{1}\cdots p_{k}\\
p_{1}^{\prime}\cdots p_{k}^{\prime}
}
}V_{p_{1}}^{\alpha_{1}}\cdots V_{p_{k}}^{\alpha_{k}}V_{p_{1}^{\prime}}^{\alpha_{1}^{\prime}}\cdots V_{p_{k}^{\prime}}^{\alpha_{k}^{\prime}}\text{Tr}_{c}[\mathcal{T}_{\tau}e^{-\beta H_{\text{bath}}}c_{p_{1}}^{\dagger}(\tau_{1})c_{p_{1}^{\prime}}(\tau_{1}^{\prime})\cdots c_{p_{k}}^{\dagger}(\tau_{k})c_{p_{k}^{\prime}}(\tau_{k}^{\prime})]\\
 & =\sum_{\substack{p_{1}\cdots p_{k}\\
p_{1}^{\prime}\cdots p_{k}^{\prime}
}
}V_{p_{1}}^{\alpha_{1}}\cdots V_{p_{k}}^{\alpha_{k}}V_{p_{1}^{\prime}}^{\alpha_{1}^{\prime}}\cdots V_{p_{k}^{\prime}}^{\alpha_{k}^{\prime}}\text{Tr}_{c}[\mathcal{T}_{\tau}e^{-\beta H_{\text{bath}}}c_{p_{1}}^{\dagger}(\tau_{1})\cdots c_{p_{k}}^{\dagger}(\tau_{k})c_{p_{1}^{\prime}}(\tau_{1}^{\prime})\cdots c_{p_{k}^{\prime}}(\tau_{k}^{\prime})]\\
 & \times(-1)^{\sum_{i=1}^{k}(i-1)},
\end{aligned}
\label{hybdef}
\end{equation}
where we define the bath partition function
\begin{equation}
Z_{\text{bath}}=\text{Tr}_c e^{-\beta H_{\text{bath}}},
\end{equation}
and the hybridization function
\begin{equation}
\triangle_{\alpha_{i}^{\prime}\alpha_{j}}(\tau_{i}^{\prime}-\tau_{j})=\sum_{p_{i}^{\prime}p_{j}}V_{p_{i}^{\prime}}^{\alpha_{i}^{\prime}*}V_{p_{j}}^{\alpha_{j}}\text{Tr}_{c}[e^{-\beta H_{\text{bath}}}c_{p_{j}}^{\dagger}(\tau_{j})c_{p_{i}^{\prime}}(\tau_{i}^{\prime})].
\end{equation}
With the above defined determinant $\det\Delta^{(\mathcal{C}_{k})}$, the partition function $Z$ can be written as
\begin{equation}
\begin{aligned}
Z & =Z_{\text{bath}}\sum_{k=0}^{\infty}\int_{0}^{\beta}d\tau_{1}\cdots\int_{\tau_{k-1}}^{\beta}d\tau_{k}\int_{0}^{\beta}d\tau_{1}^{\prime}\cdots\int_{\tau_{k-1}^{\prime}}^{\beta}d\tau_{k}^{\prime}\sum_{\substack{\alpha_{1}\cdots\alpha_{k}\\
\alpha_{1}^{\prime}\cdots\alpha_{k}^{\prime}
}
}(-1)^{\sum_{i=1}^{k}(i-1)}\det\Delta^{(\mathcal{C}_{k})}\\
 & \times\text{Tr}_{f}[\mathcal{T}_{\tau}e^{-\beta H_{\text{loc}}}f_{\alpha_{1}}(\tau_{1})\cdots f_{\alpha_{k}}(\tau_{k})f_{\alpha_{1}^{\prime}}^{\dagger}(\tau_{1}^{\prime})\cdots f_{\alpha_{k}^{\prime}}^{\dagger}(\tau_{k}^{\prime})].
\end{aligned}
\end{equation}
Contribution of configuration $\mathcal{C}_{k}$ to $Z$ can be expressed as 
\begin{equation}
\begin{aligned}
w_{Z}(\mathcal{C}_{k}) & =Z_{\text{bath}}\prod_{i=1}^{k}d\tau_{i}d\tau_{i}^{\prime}(-1)^{\sum_{i=1}^{k}(i-1)}\det\Delta^{(\mathcal{C}_{k})}\\
 & \times\text{Tr}_{f}[\mathcal{T}_{\tau}e^{-\beta H_{\text{loc}}}f_{\alpha_{1}}(\tau_{1})\cdots f_{\alpha_{k}}(\tau_{k})f_{\alpha_{1}^{\prime}}^{\dagger}(\tau_{1}^{\prime})\cdots f_{\alpha_{k}^{\prime}}^{\dagger}(\tau_{k}^{\prime})].
\end{aligned}
\end{equation}
$Z$ is just the summation over configuration space 
\begin{equation}
Z=\sum_{k}\sum_{\mathcal{C}_{k}}w_{Z}(\mathcal{C}_{k}).
\end{equation}

\section{Green's function}
In this appendix, we present a derivation of the measurement formula for $G_{\alpha\alpha'}^f(\tau)$ in Eq.~(\ref{G_KR}).
As in the partition function $Z$, we perform an expansion with respect $H_1$ as follows:
\begin{equation}
\begin{aligned}
G_{\alpha\alpha^{\prime}}(\tau,\tau^{\prime}) & \equiv G^f_{\alpha\alpha^{\prime}}(\tau,\tau^{\prime}) =-\langle\mathcal{T}_{\tau}f_{\alpha}(\tau)f_{\alpha^{\prime}}^{\dagger}(\tau^{\prime})\rangle\\
 & =-\frac{1}{Z}\text{Tr}[\mathcal{T}_{\tau}e^{-\beta H_{0}}e^{-\int_{0}^{\beta}H_{1}(\tilde{\tau})d\tilde{\tau}}f_{\alpha}(\tau)f_{\alpha^{\prime}}^{\dagger}(\tau^{\prime})]\\
 & =-\frac{1}{Z}\sum_{k=0}^{\infty}\frac{1}{(k!)^{2}}\int_{0}^{\beta}d\tau_{1}\cdots\int_{0}^{\beta}d\tau_{k}\int_{0}^{\beta}d\tau_{1}^{\prime}\cdots\int_{0}^{\beta}d\tau_{k}^{\prime}\text{Tr}[\mathcal{T}_{\tau}e^{-\beta H_{0}}V(\tau_{1})\cdots V(\tau_{k})V^{\dagger}(\tau_{1}^{\prime})\cdots V^{\dagger}(\tau_{k}^{\prime})f_{\alpha}(\tau)f_{\alpha^{\prime}}^{\dagger}(\tau^{\prime})]\\
 & =-\frac{1}{Z}\sum_{k=0}^{\infty}\frac{1}{(k!)^{2}}\int_{0}^{\beta}d\tau_{1}\cdots\int_{0}^{\beta}d\tau_{k}\int_{0}^{\beta}d\tau_{k+1}\int_{0}^{\beta}d\tau_{1}^{\prime}\cdots\int_{0}^{\beta}d\tau_{k}^{\prime}\int_{0}^{\beta}d\tau_{k+1}^{\prime}\sum_{\alpha_{k+1}}\sum_{\alpha_{k+1}^{\prime}}\delta_{\alpha_{k+1}\alpha}\delta_{\alpha_{k+1}^{\prime}\alpha^{\prime}}\\
 & \times\text{Tr}[\mathcal{T}_{\tau}e^{-\beta H_{0}}V(\tau_{1})\cdots V(\tau_{k})V^{\dagger}(\tau_{1}^{\prime})\cdots V^{\dagger}(\tau_{k}^{\prime})f_{\alpha_{k+1}}(\tau_{k+1})f_{\alpha_{k+1}^{\prime}}^{\dagger}(\tau_{k+1}^{\prime})]\delta(\tau-\tau_{k+1})\delta(\tau^{\prime}-\tau_{k+1}^{\prime}).
\end{aligned}
\end{equation}
Because $V$ and $V^{\dagger}$ are essentially bosonic, it results
in no sign by shifting $f_{\alpha_{k+1}}(\tau_{k+1})$ and $f_{\alpha_{k+1}^{\prime}}^{\dagger}(\tau_{k+1}^{\prime})$
over $V(\tau_{i})$ or $V^{\dagger}(\tau_{i}^{\prime})$. There are
$(k+1)^{2}$ different ways to move $f_{\alpha_{k+1}}(\tau_{k+1})$ to other positions among the $V(\tau_{1})\cdots V(\tau_{k})$ terms and 
$f_{\alpha_{k+1}^{\prime}}^{\dagger}(\tau_{k+1}^{\prime})$ to other positions among $V^{\dagger}(\tau_{1}^{\prime})\cdots V^{\dagger}(\tau_{k}^{\prime})$ terms. 
And since the integral is unconstrained, all these $(k+1)^2$ terms are actually equal to each other. 
For each of those $(k+1)^2$ situations , we can reindex
$(2k+2)$ operators with $f_{\alpha_{k+1}}(\tau_{k+1})$ being located at the $m$-th
location among $V(\tau_{1})\cdots V(\tau_{k})$ while $f_{\alpha_{k+1}^{\prime}}^{\dagger}(\tau_{k+1}^{\prime})$
being located at the $n$-th location among $V^{\dagger}(\tau_{1}^{\prime})\cdots V^{\dagger}(\tau_{k}^{\prime})$
\begin{equation}
\begin{aligned}
G_{\alpha\alpha^{\prime}}(\tau,\tau^{\prime}) & =-\frac{1}{Z}\sum_{k=0}^{\infty}\frac{1}{(k!)^{2}}\left[\frac{1}{(k+1)^{2}}\sum_{m,n=0}^{k+1}\right]\int_{0}^{\beta}d\tau_{1}\cdots\int_{0}^{\beta}d\tau_{k+1}\int_{0}^{\beta}d\tau_{1}^{\prime}\cdots\int_{0}^{\beta}d\tau_{k+1}^{\prime}\\
 & \times\delta(\tau-\tau_{m})\delta(\tau^{\prime}-\tau_{n}^{\prime})\sum_{\alpha_{m}}\sum_{\alpha_{n}^{\prime}}\delta_{\alpha_{m}\alpha}\delta_{\alpha_{n}^{\prime}\alpha^{\prime}}\\
 & \text{Tr}[\mathcal{T}_{\tau}e^{-\beta H_{0}}V(\tau_{1})\cdots V(\tau_{m-1})f_{\alpha_{m}}(\tau_{m})V(\tau_{m+1})\cdots V(\tau_{k+1})\\
 & \times V^{\dagger}(\tau_{1}^{\prime})\cdots V^{\dagger}(\tau_{n-1}^{\prime})f_{\alpha_{n}^{\prime}}^{\dagger}(\tau_{n}^{\prime})V^{\dagger}(\tau_{n+1}^{\prime})\cdots V^{\dagger}(\tau_{k+1}^{\prime})].
\end{aligned}
\end{equation}
Changing the unconstrained integrals to the constrained ones and plugging
in explicit forms of $V$ and $V^{\dagger}$ into Green's function,
we obtain
\begin{equation}
\begin{aligned}
G_{\alpha\alpha^{\prime}}(\tau,\tau^{\prime}) & =-\frac{1}{Z}\sum_{k=0}^{\infty}\sum_{m,n=0}^{k+1}\int_{0}^{\beta}d\tau_{1}\cdots\int_{\tau_{k}}^{\beta}d\tau_{k+1}\int_{0}^{\beta}d\tau_{1}^{\prime}\cdots\int_{\tau_{k}^{\prime}}^{\beta}d\tau_{k+1}^{\prime}\delta(\tau-\tau_{m})\delta(\tau^{\prime}-\tau_{n}^{\prime}).\\
 & \times\sum_{\substack{\alpha_{1}\cdots\alpha_{k+1}\\
\alpha_{1}^{\prime}\cdots\alpha_{k+1}^{\prime}
}
}\delta_{\alpha_{m}\alpha}\delta_{\alpha_{n}^{\prime}\alpha^{\prime}}\sum_{\substack{p_{1}\cdots p_{m-1}p_{m+1}\cdots p_{k}\\
p_{1}^{\prime}\cdots p_{n-1}^{\prime}p_{n+1}^{\prime}p_{k}^{\prime}
}
}V_{p_{1}}^{\alpha_{1}}\cdots V_{p_{m-1}}^{\alpha_{m-1}}V_{p_{m+1}}^{\alpha_{m+1}}\cdots V_{p_{k+1}}^{\alpha_{k+1}}V_{p_{1}^{\prime}}^{\alpha_{1}^{\prime}}\cdots V_{p_{n-1}^{\prime}}^{\alpha_{n-1}^{\prime}}V_{p_{n+1}^{\prime}}^{\alpha_{n+1}^{\prime}}\cdots V_{p_{k+1}^{\prime}}^{\alpha_{k+1}^{\prime}}\text{Tr}[\mathcal{T}_{\tau}e^{-\beta H_{0}}\\
 & \times c_{p_{1}}^{\dagger}(\tau_{1})f_{\alpha_{1}}(\tau_{1})\cdots c_{p_{m-1}}^{\dagger}(\tau_{m-1})f_{\alpha_{m-1}}(\tau_{m-1})f_{\alpha_{m}}(\tau_{m})c_{p_{m+1}}^{\dagger}(\tau_{m+1})f_{\alpha_{m+1}}(\tau_{m+1})\cdots c_{p_{k+1}}^{\dagger}(\tau_{k+1})f_{\alpha_{k+1}}(\tau_{k+1})\\
 & \times f_{\alpha_{1}^{\prime}}^{\dagger}(\tau_{1}^{\prime})c_{p_{1}^{\prime}}(\tau_{1}^{\prime})\cdots f_{\alpha_{n-1}^{\prime}}^{\dagger}(\tau_{n-1}^{\prime})c_{p_{n-1}^{\prime}}(\tau_{n-1}^{\prime})f_{\alpha_{n}^{\prime}}^{\dagger}(\tau_{n}^{\prime})f_{\alpha_{n+1}^{\prime}}^{\dagger}(\tau_{n+1}^{\prime})c_{p_{n+1}^{\prime}}(\tau_{n+1}^{\prime})\cdots f_{\alpha_{k+1}^{\prime}}^{\dagger}(\tau_{k+1}^{\prime})c_{p_{k+1}^{\prime}}(\tau_{k+1}^{\prime})]
\end{aligned}
\end{equation}
If we shift all conduction electron's operators to the left side and
then separate bath and impurity operators, there will be a sign, $s$,
relating to such manipulation as 
\begin{equation}
\begin{aligned}
s & =(-1)^{\sum_{i=1}^{m-1}(i-1)}\cdot(-1)^{\sum_{i=1}^{k+1-m}(i-1+m)}\cdot(-1)^{\sum_{i=1}^{n-1}(i+k+1)}\cdot(-1)^{\sum_{i=1}^{k+1-n}(i+k+1+n)}\\
 & =(-1)^{2+3k+2k^{2}-m-n}=(-1)^{k+m+n}.
\end{aligned}
\end{equation}
As a result, 
\begin{equation}
\begin{aligned}
G_{\alpha\alpha^{\prime}}(\tau,\tau^{\prime}) & =-\frac{1}{Z}\sum_{k=0}^{\infty}\sum_{m,n=0}^{k+1}\int_{0}^{\beta}d\tau_{1}\cdots\int_{\tau_{k}}^{\beta}d\tau_{k+1}\int_{0}^{\beta}d\tau_{1}^{\prime}\cdots\int_{\tau_{k}^{\prime}}^{\beta}d\tau_{k+1}^{\prime}\delta(\tau-\tau_{m})\delta(\tau^{\prime}-\tau_{n}^{\prime})\cdot(-1)^{k+m+n}\\
 & \times \sum_{\substack{\alpha_{1}\cdots\alpha_{k+1}\\
\alpha_{1}^{\prime}\cdots\alpha_{k+1}^{\prime}
}
}\delta_{\alpha_{m}\alpha}\delta_{\alpha_{n}^{\prime}\alpha^{\prime}}\sum_{\substack{p_{1}\cdots p_{m-1}p_{m+1}\cdots p_{k+1}\\
p_{1}^{\prime}\cdots p_{n-1}^{\prime}p_{n+1}^{\prime}p_{k+1}^{\prime}
}
}V_{p_{1}}^{\alpha_{1}}\cdots V_{p_{m-1}}^{\alpha_{m-1}}V_{p_{m+1}}^{\alpha_{m+1}}\cdots V_{p_{k+1}}^{\alpha_{k+1}}V_{p_{1}^{\prime}}^{\alpha_{1}^{\prime}}\cdots V_{p_{n-1}^{\prime}}^{\alpha_{n-1}^{\prime}}V_{p_{n+1}^{\prime}}^{\alpha_{n+1}^{\prime}}\cdots V_{p_{k+1}^{\prime}}^{\alpha_{k+1}^{\prime}}\\
 & \times\text{Tr}_{c}[\mathcal{T}_{\tau}e^{-\beta H_{bath}}c_{p_{1}}^{\dagger}(\tau_{1})\cdots c_{p_{m-1}}^{\dagger}(\tau_{m-1})c_{p_{m+1}}^{\dagger}(\tau_{m+1})\cdots c_{p_{k+1}}^{\dagger}(\tau_{k+1})\\
 & \times\hfill\quad c_{p_{1}^{\prime}}(\tau_{1}^{\prime})\cdots c_{p_{n-1}^{\prime}}(\tau_{n-1}^{\prime})c_{p_{n+1}^{\prime}}(\tau_{n+1}^{\prime})\cdots c_{p_{k+1}^{\prime}}(\tau_{k+1}^{\prime})]\\
 & \times\text{Tr}_{f}[\mathcal{T}_{\tau}e^{-\beta H_{loc}}f_{\alpha_{1}}(\tau_{1})\cdots f_{\alpha_{k+1}}(\tau_{k+1})f_{\alpha_{1}^{\prime}}^{\dagger}(\tau_{1}^{\prime})\cdots f_{\alpha_{k+1}^{\prime}}^{\dagger}(\tau_{k+1}^{\prime})].
\end{aligned}
\end{equation}
According to Eq.~(\ref{hybdef}), we have
\begin{equation}
\begin{aligned}
 &\frac{1}{Z_{\text{bath}}}\sum_{\substack{p_{1}\cdots p_{m-1}p_{m+1}\cdots p_{k+1}\\
p_{1}^{\prime}\cdots p_{n-1}^{\prime}p_{n+1}^{\prime}p_{k+1}^{\prime}
}
}V_{p_{1}}^{\alpha_{1}}\cdots V_{p_{m-1}}^{\alpha_{m-1}}V_{p_{m+1}}^{\alpha_{m+1}}\cdots V_{p_{k+1}}^{\alpha_{k+1}}V_{p_{1}^{\prime}}^{\alpha_{1}^{\prime}}\cdots V_{p_{n-1}^{\prime}}^{\alpha_{n-1}^{\prime}}V_{p_{n+1}^{\prime}}^{\alpha_{n+1}^{\prime}}\cdots V_{p_{k+1}^{\prime}}^{\alpha_{k+1}^{\prime}}\text{Tr}_{c}[\mathcal{T}_{\tau}e^{-\beta H_{bath}}\\
\times & c_{p_{1}}^{\dagger}(\tau_{1})\cdots c_{p_{m-1}}^{\dagger}(\tau_{m-1})c_{p_{m+1}}^{\dagger}(\tau_{m+1})\cdots c_{p_{k+1}}^{\dagger}(\tau_{k+1})c_{p_{1}^{\prime}}(\tau_{1}^{\prime})\cdots c_{p_{n-1}^{\prime}}(\tau_{n-1}^{\prime})c_{p_{n+1}^{\prime}}(\tau_{n+1}^{\prime})\cdots c_{p_{k+1}^{\prime}}(\tau_{k+1}^{\prime})]\\
= & (-1)^{\sum_{i=1}^{k}(i-1)}\Delta_{(\bcancel{n,m})}^{(\mathcal{C}_{k+1})}
\end{aligned}
\end{equation}
where $\Delta_{(\bcancel{n,m})}^{(\mathcal{C}_{k+1})}$ is obtained
from $\Delta^{(\mathcal{C}_{k+1})}$ by deleting $n$-th row and $m$-th
colum. Green's function now reads
\begin{equation}
\begin{aligned}
G_{\alpha\alpha^{\prime}}(\tau,\tau^{\prime}) & =-\frac{Z_{\text{bath}}}{Z}\sum_{k=0}^{\infty}\sum_{m,n=0}^{k+1}\int_{0}^{\beta}d\tau_{1}\cdots\int_{\tau_{k}}^{\beta}d\tau_{k+1}\int_{0}^{\beta}d\tau_{1}^{\prime}\cdots\int_{\tau_{k}^{\prime}}^{\beta}d\tau_{k+1}^{\prime}\delta(\tau-\tau_{m})\delta(\tau^{\prime}-\tau_{n}^{\prime})\\
 & \times\sum_{\substack{\alpha_{1}\cdots\alpha_{k+1}\\
\alpha_{1}^{\prime}\cdots\alpha_{k+1}^{\prime}
}
}\delta_{\alpha_{m}\alpha}\delta_{\alpha_{n}^{\prime}\alpha^{\prime}}\det\Delta_{(\bcancel{n,m})}^{(\mathcal{C}_{k+1})}\cdot(-1)^{\sum_{i=1}^{k}(i-1)}\cdot(-1)^{k+m+n}\\
 & \times\text{Tr}_{f}[\mathcal{T}_{\tau}e^{-\beta H_{loc}}f_{\alpha_{1}}(\tau_{1})\cdots f_{\alpha_{k+1}}(\tau_{k+1})f_{\alpha_{1}^{\prime}}^{\dagger}(\tau_{1}^{\prime})\cdots f_{\alpha_{k+1}^{\prime}}^{\dagger}(\tau_{k+1}^{\prime})].
\end{aligned}
\end{equation}
Configuration spaces of Green's function at $k$-th order can be represented
by those of partition function at $k+1$ order, $\mathcal{C}_{k}^{G}\equiv\mathcal{C}_{k+1}$,
which contribute to Green's function with the weight
\begin{equation}
\begin{aligned}
w_{G}(\mathcal{C}_{k+1})= & -Z_{\text{bath}}\prod_{i=1}^{k+1}d\tau_{i}d\tau_{i}^{\prime}\sum_{m,n=0}^{k+1}\delta_{\alpha_{m}\alpha}\delta_{\alpha_{n}^{\prime}\alpha^{\prime}}\det\Delta_{(\bcancel{n,m})}^{(\mathcal{C}_{k+1})}\cdot(-1)^{\sum_{i=1}^{k}(i-1)}\cdot(-1)^{k+m+n}\\
 & \times\delta(\tau-\tau_{m})\delta(\tau^{\prime}-\tau_{n}^{\prime})\text{Tr}_{f}[\mathcal{T}_{\tau}e^{-\beta H_{loc}}f_{\alpha_{1}}(\tau_{1})\cdots f_{\alpha_{k+1}}(\tau_{k+1})f_{\alpha_{1}^{\prime}}^{\dagger}(\tau_{1}^{\prime})\cdots f_{\alpha_{k+1}^{\prime}}^{\dagger}(\tau_{k+1}^{\prime})].
\end{aligned}
\end{equation}
For the sake of convenience, we here give the contribution of configuration $\mathcal{C}_{k+1}$ to $Z$ 
\begin{equation}
\begin{aligned}
w_{Z}(\mathcal{C}_{k+1}) & =Z_{\text{bath}}\prod_{i=1}^{k+1}d\tau_{i}d\tau_{i}^{\prime}(-1)^{\sum_{i=1}^{k+1}(i-1)}\det\Delta^{(\mathcal{C}_{k+1})}\\
 & \times\text{Tr}_{f}[\mathcal{T}_{\tau}e^{-\beta H_{loc}}f_{\alpha_{1}}(\tau_{1})\cdots f_{\alpha_{k+1}}(\tau_{k+1})f_{\alpha_{1}^{\prime}}^{\dagger}(\tau_{1}^{\prime})\cdots f_{\alpha_{k+1}^{\prime}}^{\dagger}(\tau_{k+1}^{\prime})].
\end{aligned}
\end{equation}
The measurement of Green's function is
\begin{equation}
\begin{aligned}
G_{\alpha\alpha^{\prime}}(\tau,\tau^{\prime}) & =\frac{1}{Z}\sum_{k=0}^{\infty}\sum_{\mathcal{C}_{k+1}}\frac{w_{G}(\mathcal{C}_{k+1})}{w_{Z}(\mathcal{C}_{k+1})}w_{Z}(\mathcal{C}_{k+1})=\langle\frac{w_{G}(\mathcal{C}_{k+1})}{w_{Z}(\mathcal{C}_{k+1})}\rangle_{Z}\\
 & =\langle-\sum_{m,n=0}^{k+1}\delta_{\alpha_{m}\alpha}\delta_{\alpha_{n}^{\prime}\alpha^{\prime}}\delta(\tau-\tau_{m})\delta(\tau^{\prime}-\tau_{n}^{\prime})\cdot\frac{(-1)^{m+n}\det\Delta_{(\bcancel{n,m})}^{(\mathcal{C}_{k+1})}}{\det\Delta^{(\mathcal{C}_{k+1})}}\rangle_{Z}\\
 & =\langle-\frac{1}{\beta}\sum_{m,n=0}^{k+1}\delta_{\alpha_{m}\alpha}\delta_{\alpha_{n}^{\prime}\alpha^{\prime}}\delta(\tau-\tau^{\prime},\tau_{m}-\tau_{n}^{\prime})\mathcal{M}_{mn}^{(\mathcal{C}_{k+1})}\rangle_{Z}.
\end{aligned}
\label{Gf_W2W}
\end{equation}
where $\mathcal{M}^{(\mathcal{C}_{k+1})}=[\Delta^{(\mathcal{C}_{k+1})}]^{-1}$.
The arguments, $\tau$ and $\tau^{\prime}$, are in $[0,\beta]$ as
a priori. Actually, $G_{\alpha\alpha^{\prime}}(\tau,\tau^{\prime})$
is a $\beta$-antiperiodic function of $\tau-\tau^{\prime}$. To restore
$\beta$-antiperiodicity, $\delta(\tau)$ is replaced by Dirac comb
defined as 
\begin{equation}
\delta^{-}(\tau)\equiv\sum_{l\in\mathbb{Z}}(-1)^{l}\delta(\tau-l\beta).
\end{equation}
Using the translational invariance 
\begin{equation}
G_{\alpha\alpha^{\prime}}(\tau-\tau^{\prime})=\frac{1}{\beta}\int_{0}^{\beta}dsG_{\alpha\alpha^{\prime}}(\tau+s,\tau^{\prime}+s)
\end{equation}
we finally get the measurement formula for Green's function
\begin{equation}
G_{\alpha\alpha^{\prime}}(\tau-\tau^{\prime})=-\langle\frac{1}{\beta}\sum_{m,n=1}^{k+1}\delta_{\alpha_{m},\alpha}\delta_{\alpha_{n}^{\prime},\alpha^{\prime}}\delta^{-}[\tau-\tau^{\prime}-(\tau_{m}-\tau_{n}^{\prime})]\mathcal{M}_{nm}^{(\mathcal{C}_{k+1})}\rangle_{Z}.
\end{equation}
Replacing $k+1$ by $k$ ( since we typically call ``current'' configuration
as $\mathcal{C}_{k}$), we have 
\begin{equation}
G_{\alpha\alpha^{\prime}}(\tau-\tau^{\prime})=-\langle\frac{1}{\beta}\sum_{m,n=1}^{k}\delta_{\alpha_{m},\alpha}\delta_{\alpha_{n}^{\prime},\alpha^{\prime}}\delta^{-}[\tau-\tau^{\prime}-(\tau_{m}-\tau_{n}^{\prime})]\mathcal{M}_{nm}^{(\mathcal{C}_{k})}\rangle_{Z}.
\label{Gf_CTHYB}
\end{equation}
Eq.~(\ref{Gf_CTHYB}) is just the measurement formula of Green's function for CT-HYB. 

We emphasize that the local trace, $\text{Tr}_{f}[\cdots]$, is completely
canceled in $\langle\frac{w_{G}(\mathcal{C}_{k+1})}{w_{Z}(\mathcal{C}_{k+1})}\rangle_{Z}$
of Eq.~(\ref{Gf_W2W}), which means that the measurement formula of Green's function
in CT-X has the same form to that in CT-HYB since approximations in
CT-X are only made to the local trace. However, one has to reinterpret
the configuration spaces: under the approximations made in CT-X, $2k$
separated operators form number of $k$ pair-operators as encoded
in the definition of $X$ matrices, Eq.~(\ref{xmatp1}-\ref{xmatm1}). As a result, summation
over separated creation ($\sum_{n}$) and annihilation ($\sum_{m}$)
operators in CT-HYB is reinterpreted as summation over pair-operators:
$f_{\alpha_{m}}$ is from the $m$-th $X$-matrix located at $\tau_{m}$
while $f_{\alpha^\prime_{n}}^{\dagger}$ is from the $n$-th $X$-matrix
located at $\tau_{n}\equiv\tau_{n}^{\prime}$. Last but not the least,
we have to discard summands with $m=n$ which are from high energy
processes. Finally, we arrive at the measurement formula of Green's
function in CT-X
\begin{equation}
G_{\alpha\alpha^{\prime}}(\tau)=-\langle\frac{1}{\beta}\sum_{m,n=1}^{\,k\,\prime}\delta_{\alpha_{m},\alpha}\delta_{\alpha_{n}^{\prime},\alpha^{\prime}}\delta^{-}[\tau-(\tau_{m}-\tau_{n})]\mathcal{M}_{nm}^{(\mathcal{C}_{k})}\rangle_{Z}.
\end{equation}

\end{widetext}
\clearpage
\bibliography{ctx_ref}
\end{document}